\documentclass{article}

\usepackage{PRIMEarxiv}

\usepackage[utf8]{inputenc} 
\usepackage[T1]{fontenc}    
\usepackage{hyperref}       
\usepackage{url}            
\usepackage{booktabs}       
\usepackage{amsfonts}       
\usepackage{nicefrac}       
\usepackage{microtype}      
\usepackage{fancyhdr}       
\usepackage{graphicx}       
\usepackage{tikz}
\usetikzlibrary{arrows.meta,calc,patterns}
\usepackage[table]{xcolor}
\usepackage{amsmath}
\usepackage{algorithm}
\usepackage[noend]{algorithmic}
\usepackage{setspace}

\usepackage{siunitx}
\DeclareMathOperator*{\argmax}{argmax}

\newcommand{\normal}[1]{\mathcal{N}\!\left(#1\right)\!}
\newcommand{\rbracket}[1]{\mathinner{\!\left(#1\right)\!}}
\newcommand{\sbracket}[1]{\mathinner{\!\left[#1\right]\!}}
\newcommand{\cbracket}[1]{\mathinner{\!\left\{#1\right\}\!}}

\newcommand*{\defeq}{\mathrel{\vcenter{\baselineskip0.5ex \lineskiplimit0pt \hbox{\scriptsize.}\hbox{\scriptsize.}}}=}

\newcommand{\xb}{\boldsymbol{x}}

\newcommand{\yb}{\boldsymbol{y}}

\newcommand{\epsb}{\boldsymbol{\epsilon}}

\newcommand{\thetab}{\boldsymbol{\theta}}
\newcommand{\mub}{\boldsymbol{\mu}}

\newcommand{\xib}{\boldsymbol{\xi}}

\newcommand{\E}{\mathbb{E}}

\newcommand{\nnew}{N_{\mathrm{new}}}
\newcommand{\EIG}{\mathrm{EIG}}
\newcommand{\IG}{\mathrm{IG}}
\newcommand{\model}{\mathcal{M}}

\newenvironment{eq}{\begin{equation}\begin{aligned}}{\end{aligned}\end{equation}\ignorespacesafterend}

\makeatletter
\newcommand\thefontsize[1]{{#1 The current font size is: \f@size pt\par}}
\makeatother

\usepackage[style=numeric, sorting=none, giveninits=true, isbn=false,eprint=false, minnames=15, maxnames=20, backend=biber]{biblatex}
\bibliography{main}  
\AtEveryBibitem{\clearfield{note}}
\AtEveryBibitem{\clearlist{language}}
\AtEveryBibitem{\clearfield{urlyear}}
\AtEveryBibitem{\ifentrytype{misc}{%
  }{%
    \clearfield{url}%
  }%
}

\definecolor{cblue}{rgb}{0.050383, 0.029803, 0.527975}
\definecolor{cred}{rgb}{0.881443, 0.392529, 0.383229}
\definecolor{cyellow}{rgb}{0.988648, 0.809579, 0.145357}

\hypersetup{
     linkbordercolor=cblue,
     citebordercolor=cred,   
     urlbordercolor=cyellow
}

\title{Efficient Bayesian Optimal Experimental Design\\for Expensive Computational Models\\over Finite Design Sets}

\author{Maximilian Dinkel\thanks{corresponding author: \texttt{maximilian.dinkel@tum.de}} $^{, a}$, Dragos C. Ana$^{a,b}$, Benedikt Goderbauer$^{a}$ and Wolfgang A. Wall$^{a,b,c}$ \\
[1em]
$^{a}$ Institute for Computational Mechanics \\
Technical University of Munich \\
Boltzmannstr. 15, 85748 Garching, Germany\\
[1em]
$^{b}$ TUMint.Energy Research\\
Lichtenbergstr. 4, 85748 Garching, Germany\\
[1em]
$^{c}$ Munich Data Science Institute\\
Technical University of Munich \\
Walther-Von-Dyck Str. 10, 85748 Garching, Germany\\
}

\begin{document}
\maketitle
\delimitershortfall=-1pt
\setlength{\parindent}{0pt}

\begin{abstract}
Bayesian calibration is a powerful framework for identifying parameters in complex and large-scale computational models. However, when there is insufficient or poorly suited data for the calibration process, significant uncertainty about the identified parameters remains. This uncertainty can hinder effective decision-making and understanding of the system. In many applications, the experiment that generates the data can be influenced by design variables, such as sensor placements, loading conditions, or test configurations. Often, however, these design variables can not be chosen arbitrarily but only from a finite set of possible experimental designs. \\
We propose an adaptive algorithm for Bayesian optimal experimental design over a finite design set, specifically tailored for applications involving expensive computational models. The method integrates an accelerated nested Monte Carlo estimator that reuses parameter samples to reduce model evaluations. Additionally, it employs common random numbers and Rao--Blackwellization to reduce variance in pairwise expected information gain comparisons. To support decision-making within the algorithm, we estimate the probability that one design outperforms another using bootstrap sampling. Starting with small sample sizes, the algorithm iteratively eliminates inferior designs based on these probabilistic comparisons, allocating additional computational effort only to promising candidates until a single design remains.
The resulting approach achieves high reliability at substantially reduced computational cost, making it well-suited for large-scale engineering applications.

\end{abstract}

\keywords{Bayesian Experimental Design, Finite Design Sets, Adaptive Elimination, Nested Monte Carlo, Rao--Blackwellization, Common Random Numbers, Bootstrap}

\section{Introduction}
\label{Introduction}
Computational models are indispensable across engineering and more or less all the sciences for insight, prediction, design, and decision support. When model parameters $\thetab$ are uncertain, Bayesian calibration provides a principled way to combine prior knowledge with experimental or observational data to obtain posterior distributions over the parameters \cite{kennedy_bayesian_2001}. The utility of Bayesian calibration, however, hinges on the informativeness of the data \cite{haeusel_multi-physics-enhanced_2026}. In many settings, the data-generating process (e.g. a physical experiment) can be influenced by controllable design variables, such as sensor placement, loading conditions, or test configurations. This motivates Bayesian optimal experimental design (BOED), in which one seeks the design that maximizes the expected information gain (EIG), i.e., the expected reduction of entropy from the prior to the posterior distribution \cite{lindley_measure_1956, lindley_bayesian_1972, bernardo_expected_1979, chaloner_bayesian_1995, sebastiani_maximum_2000, ryan_review_2016, rainforth_modern_2024, huan_optimal_2024}. \\
By contrast, frequentist experimental design approaches are typically based on the Fisher information matrix \cite{atkinson_optimum_2007, rainforth_modern_2024, huan_optimal_2024}, which depends on the unknown true parameters $\thetab$. In general settings, this dependence necessitates either model approximations \cite{hughes-olivier_optimal_1998}, the use of point estimates or representative parameter values for $\thetab$ \cite{atkinson_optimum_2007, chen_minimax_2015}, or averaging over a prior distribution \cite{atkinson_optimum_2007,chaloner_optimal_1989,chaloner_bayesian_1995,firth_parameter_1997}. By integrating over the prior when evaluating the EIG, BOED naturally accounts for parameter uncertainty \cite{rainforth_modern_2024}. \\
Despite its conceptual appeal, computing and optimizing the EIG is challenging because it involves high-dimensional expectations over both parameters and observations, often requiring repeated evaluations of expensive computational models.
A common approach is to leverage Laplace approximations within 
nested estimation schemes, including both fully deterministic 
nested Laplace methods and hybrids where Laplace approximations 
are used to construct importance sampling proposals \cite{lewi_sequential_2009, long_fast_2013, long_fast_2015, long_laplace_2015, long_multimodal_2022, ryan_fully_2015, beck_fast_2018}, which are computationally efficient but often rely on local Gaussian assumptions \cite{rainforth_modern_2024}. Consequently, the resulting EIG estimates are generally biased. Moreover, Laplace methods require access to derivatives of the log-likelihood to compute Hessians, which is often infeasible for legacy computational models, and approximating second-order derivatives via finite differences can be numerically challenging in high-dimensional spaces \cite{fowkes_approximating_2025}. \\
An alternative is nested Monte Carlo (NMC) \cite{ryan_towards_2014, ryan_estimating_2003, rainforth_modern_2024}, which is asymptotically consistent but suffers from high cost and slow convergence, with complexity scaling as  $C=NM$ (with $N$ outer samples and $M$ inner samples) and mean squared error rates of $\mathcal{O}(C^{-2/3})$ under optimal allocation \cite{rainforth_nesting_2018}. Moreover, NMC estimators are biased for finite inner-loop sample sizes due to the nonlinear log transformation of the inner estimators \cite{rainforth_nesting_2018}.
Several strategies have emerged to address these limitations. More generally, importance sampling improves NMC by drawing inner samples from tuned proposals, reducing variance and bias \cite{foster_variational_2019}. Debiasing schemes, such as multilevel Monte Carlo estimators, can eliminate bias entirely by expressing EIG as a telescoping sum and thereby achieving convergence rates of $\mathcal{O}(C^{-1/2})$ that are comparable to those of standard Monte Carlo \cite{giles_multilevel_2008, goda_unbiased_2022, rhee_unbiased_2015}. \\
Besides sampling-based methods, variational approaches have shown promising results by utilizing different bounds on the EIG, including upper bounds based on evidence approximations and lower bounds using amortized inference networks, each offering distinct trade-offs between bias, variance, and computational cost \cite{foster_variational_2019, rainforth_modern_2024, dong_variational_2025}. \\
Ultimately, the goal of BOED is not to estimate EIG, but to optimize it over the design space. Classical optimization strategies, such as enumeration and Bayesian optimization \cite{foster_variational_2019, snoek_practical_2012}, often wrap around EIG estimators and require numerous evaluations, making them costly, especially in high-dimensional settings \cite{rainforth_modern_2024}. Stochastic gradient methods offer improved scalability by directly estimating EIG derivatives \cite{goda_unbiased_2022, ao_estimating_2024}. However, they require gradients of the computational model with respect to the design  parameters, which is not viable for black-box simulators. \\
In many real-world applications, the space of feasible experimental designs is naturally discrete and finite. Examples include selecting among a predefined set of sensor locations, load configurations, or test protocols. In such cases, gradient-based optimization is often not applicable, and exhaustive evaluation of all candidates is computationally prohibitive due to the cost of forward model evaluations. This motivates the development of adaptive strategies that can efficiently search over discrete design sets by eliminating suboptimal candidates early, while preserving  reliability. \\
In this work, we propose an adaptive elimination algorithm for BOED over a finite design set, tailored to expensive computational models. Our approach combines an efficient NMC estimator that reuses outer-loop parameter samples to approximate the evidence, thereby drastically reducing forward model evaluations \cite{huan_accelerated_2010, huan_simulation-based_2013}. It further leverages common random numbers \cite{glasserman_guidelines_1992, rubinstein_simulation_2016} across designs and Rao--Blackwellization \cite{robert_monte_2004} to reduce the variance of pairwise EIG differences. Finally, it employs bootstrap sampling \cite{efron_introduction_1994} to estimate the probability that one design outperforms another, enabling robust elimination decisions without distributional assumptions. Starting from small sample sizes, the algorithm iteratively identifies the current best design under the NMC estimator, tests competing designs for inferiority using bootstrap on paired differences, and increases the sample size only for surviving candidates. This process continues until a single design remains, yielding high computational efficiency and reliability.

\section{Methodology}
\label{chap:methodology}

Consider we want to calibrate the unknown model parameters $\thetab$ of a deterministic computational model $\model$ based on the outcome of an experiment, denoted by $\yb$. In many scenarios, the experiment and consequently the outcome $\yb$ can be influenced by some design parameters $\xib$. The goal is to select the design parameters $\xib$, such that the uncertainty about the parameters $\thetab$ is minimized after incorporating the knowledge about the outcome of the experiment $\yb$. The outcome of the experiment $\yb$ can be generated artificially as the output of the computational model $\model$ and perturbing it using a noise model $f$ with a random variable $\epsb \sim p(\epsb)$:
\begin{eq}
    \yb = f\rbracket{\model(\thetab, \xib), \thetab, \xib, \epsb}.
    \label{eq:obs}
\end{eq}
We assume that evaluations of the computational model $\model$ are computationally expensive, as it may represent a complex simulation (e.g., a finite element or multiphysics model), whereas sampling from the noise model $f$ incurs negligible computational cost in comparison.
A popular choice for such a noise model could, for example, be an additive noise model of the form $f=\model(\thetab, \xib) + \epsb$. Nevertheless, the definition in \eqref{eq:obs} is very general, and we require no continuity with respect to $\xib$, which allows for a wide variety of possible noise models and engineering applications.

\subsection{Bayesian Optimal Experimental Design}

In the Bayesian framework, our beliefs about the unknown model parameters $\thetab$ are updated after observing data $\yb$ through Bayes' theorem. The posterior distribution is given by:
\begin{eq}
    p(\thetab \mid \yb, \xib) = \frac{p(\yb \mid \thetab, \xib) \, p(\thetab)}{p(\yb \mid \xib)},
\end{eq}
where $p(\thetab)$ is the prior, $p(\yb \mid \thetab, \xib)$ is the likelihood, and $p(\yb \mid \xib) = \int p(\yb \mid \thetab, \xib) p(\thetab) d\thetab$ is the marginal likelihood or evidence. The prior distribution $p(\thetab)$ is independent of the choice of the experimental design, and the effect of the design choice is incorporated in the likelihood $p(\yb \mid \thetab, \xib)$. \\
To quantify the informativeness of a design $\xib$, we can compare the uncertainty about the unknown parameters $\thetab$, before and after incorporating the data $\yb$. Commonly, the Shannon entropy \cite{shannon_mathematical_1948}:
\begin{eq}
    \mathrm{H}[p(\thetab)] = - \int p(\thetab) \log p(\thetab)
\end{eq}
is used to measure the uncertainty of a random variable. This gives rise to the information gain (IG), defined as the reduction in Shannon entropy from the prior to the posterior distribution \cite{lindley_measure_1956}:
\begin{eq}
    \IG(\xib, \yb) = \mathrm{H}[p(\thetab)] - \mathrm{H}[p(\thetab \mid \yb, \xib)].
\end{eq}
Since the outcome $\yb$ depends on the design $\xib$ and is unknown prior to conducting the experiment, we cannot directly optimize the information gain $\IG(\xib, \yb)$. Instead, we maximize its expectation over all possible outcomes, leading to the expected information gain $\EIG$ \cite{bernardo_expected_1979, lindley_measure_1956} (see also Appendix~\ref{appendix:eig}):
\begin{eq}
    \EIG(\xib) 
    &= \E_{p(\yb \mid \xib)} \sbracket{ \IG(\xib, \yb) } \\
    &=  \int p(\yb \mid \thetab, \xib) p(\thetab) \underbrace{\sbracket{\log p(\yb \mid \thetab, \xib) - \log p(\yb \mid \xib)}}_{=:\mathrm{PMI(\yb, \thetab \mid \xib)}} d\thetab d\yb, 
\end{eq}
where $\mathrm{PMI}(\yb,\thetab \mid \xib)$ denotes the pointwise mutual information between the observations $\yb$ and the parameters $\thetab$ given the design $\xib$. Hence, the EIG equals the mutual information between $\yb$ and $\thetab$ given $\xib$ \cite{rainforth_modern_2024}.
Equivalently, the expected information gain can be expressed as the expected Kullback-Leibler divergence between the posterior and prior \cite{rainforth_modern_2024}.
The objective in BOED can be formalized as finding the optimal experimental design $\xib^*$ that maximizes the EIG:
\begin{eq}
    \xib^* = \argmax_{\xib \in \Xi} \EIG(\xib).
\end{eq}
However, maximizing the EIG, let alone approximating the EIG, poses a major challenge, since computing the evidence $p(\yb \mid \xib)$ and the outer integral are intractable. In this work, we consider  $\Xi$ to be a finite, unordered set. Such discrete design spaces commonly arise in practice, for instance, when selecting from a fixed library of materials, boundary conditions in a test rig, or test protocols. So the optimization problem reduces to selecting the most informative element of a finite set.

\subsection{Monte Carlo Integration}

A common approach to estimate the EIG is based on NMC integration \cite{rainforth_nesting_2018}:
\begin{eq}
    \label{eqn:nmc}
    \widehat{\EIG}_{\mathrm{NMC}}(\xib) =\frac{1}{N}\sum_{i=1}^{N}  \log p(\yb_i \mid \thetab_i, \xib) - \log \sbracket{\frac{1}{M}\sum_{j=1}^{M} p(\yb_i \mid \thetab_j', \xib) },
\end{eq}
where $\yb_i \sim p(\yb \mid \thetab_i, \xib), \, \thetab_i \sim p(\thetab), \, \thetab_j' \sim p(\thetab)$. 
However, the NMC estimator is computationally expensive due to the double-loop structure and repeated evaluations of the likelihood function. Moreover, it introduces a positive bias in the estimation of the expected information gain, particularly when the number of inner samples $M$ is small. Despite this bias, the estimator remains consistent, meaning that the approximation converges to the true EIG as both $N$ and $M$ tend to infinity \cite{rainforth_nesting_2018}. \\
To reduce the computational cost of NMC estimation, one can reuse the samples from the outer loop for the inner loop. Specifically, instead of drawing a separate set of samples $\{\thetab_j'\}_{i=1}^{M}$ for each inner estimation, the same set $\Theta =\{\thetab_i\}_{i=1}^{N}$ used to generate $\yb_i$ can be repurposed to approximate the marginal likelihood as proposed in \cite{huan_accelerated_2010, huan_simulation-based_2013}. Here, we refer to this estimator as the Reuse NMC (RNMC) estimator:
\begin{eq}
    \widehat{\EIG}_{\mathrm{RNMC}}(\xib) = \frac{1}{N}\sum_{i=1}^N  \log p(\yb_i \mid \thetab_i, \xib) - \log \sbracket{\frac{1}{N}\sum_{\substack{j=1}}^N p(\yb_i \mid \thetab_j, \xib) }, 
    \label{eqn:rnmc}
\end{eq}
where $\thetab_i \in \Theta$ and $\yb_i \sim p(\yb \mid \thetab_i, \xib)$ is generated via $\model(\thetab_i,\xib)$ as in \eqref{eq:obs}. This reuse strategy reduces the number of model evaluations by a factor of $M$, thereby significantly improving computational efficiency.
\\
Opposed to the approach in \cite{huan_accelerated_2010, huan_simulation-based_2013}, we propose to exclude the current outer loop sample $\thetab_i$ from the sample set $\Theta$ for the inner expectation. This leads to the following Leave-out RNMC (LRNMC) estimator:
\begin{eq}
    \widehat{\EIG}_{\mathrm{LRNMC}}(\xib) = \frac{1}{N}\sum_{i=1}^N  \log p(\yb_i \mid \thetab_i, \xib) - \log \sbracket{\frac{1}{N-1}\sum_{\substack{j=1 \\ j \neq i}}^N p(\yb_i \mid \thetab_j, \xib) }. 
    \label{eqn:lrnmc}
\end{eq}
This is an important detail of our approach, as the standard NMC estimator from Equation \eqref{eqn:nmc} is positively biased and the additional bias introduced by reusing the current outer-loop sample within the inner loop as in Equation~\eqref{eqn:rnmc} adds a large negative bias term to the estimator. Consequently, the overall bias appears to be negative. 
By excluding the current outer-loop sample, the LRNMC estimator removes this reuse-induced negative bias term while retaining the same fundamental bias as the standard NMC estimator.
The additional bias term arises because, when we include the current outer loop sample, we evaluate the likelihood at the sample used to generate the observations. On average, the likelihood is greater at this sample than when sampled from the prior. Due to the negative sign, the additional bias contribution is negative. We explain in the subsequent section why avoiding this additional bias term is beneficial for our proposed method. \\
Due to our assumption about the noise model in \eqref{eq:obs}, we can reformulate the EIG as
\begin{eq}
    \EIG(\xib) 
    &= \int p(\thetab) \,p(\epsb)\,\sbracket{\log p\rbracket{\yb(\thetab,\xib,\epsb)\mid \thetab,\xib} - \log p\rbracket{\yb(\thetab,\xib,\epsb)\mid \xib}} \, d\epsb\, d\thetab  \\
    &= \E_{\thetab}\sbracket{\E_{\epsb}\sbracket{\mathrm{PMI} \rbracket{\yb(\thetab,\xib,\epsb),\thetab\mid\xib}}}.
\end{eq}
For fixed $\thetab$, the only randomness in the PMI arises from the noise $\epsb$, which allows us to draw multiple noise samples without additional evaluations of the computational model $\model$. Averaging these contributions yields a Monte Carlo approximation of the Rao--Blackwellized \cite{robert_monte_2004} PMI
\begin{eq}
\mathrm{PMI}^{\mathrm{RB}}(\thetab\mid \xib) := \E_{\epsb}\sbracket{\mathrm{PMI}\rbracket{\yb(\thetab,\xib,\epsb),\thetab\mid\xib}}.
\end{eq}
We refer to the corresponding estimator of the EIG as the Accelerated NMC (ANMC) estimator:
\begin{eq}
    \label{eqn:anmc}
    \widehat{\EIG}_{\mathrm{ANMC}}(\xib)
    = \frac{1}{N}\sum_{i=1}^N \underbrace{\frac{1}{N_\epsilon}\sum_{k=1}^{N_\epsilon} \log p(\yb(\thetab_i,\xib,\epsb_k)\mid \thetab_i,\xib) - \log\sbracket{\frac{1}{N-1}\sum_{\substack{j=1 \\ j\neq i}}^N p(\yb(\thetab_i,\xib,\epsb_k)\mid \thetab_j,\xib)}}_{=:\,\widehat{\mathrm{PMI}}^{\mathrm{RB}}_i(\Theta,\xib)},
\end{eq}
where $\epsb_k \sim p(\epsb)$ and $\widehat{\mathrm{PMI}}^{\mathrm{RB}}_i(\Theta,\xib)$ is a Monte Carlo estimate of $\mathrm{PMI}^{\mathrm{RB}}(\thetab_i\mid\xib)$. This conditional averaging does not change the expected value relative to the single-noise (LRNMC) case, but it reduces the variance monotonically with $N_\epsilon$. 
As a consequence, the ANMC and LRNMC estimators exhibit the same bias behavior as the standard NMC estimator with respect to the number of inner-loop samples.
Setting $N_\epsilon=1$ recovers the LRNMC estimator in \eqref{eqn:lrnmc}. We set $N_\epsilon=100$ for ANMC across all numerical examples, as we observed empirically that larger values of $N_\epsilon$ typically yield only marginal improvements in estimator variance.

\subsection{Comparing Expected Information Gains}
\begin{figure}[h!]
    \centering
    \includegraphics[width=\linewidth]{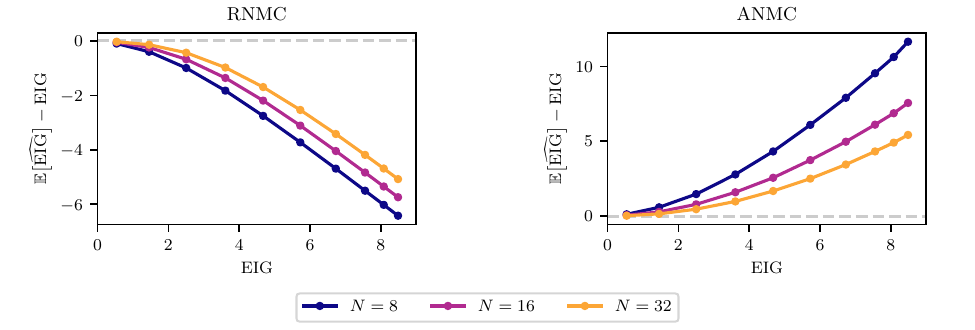}
    \caption{Bias of the RNMC and ANMC estimator of the EIG for a problem with lognormal prior and likelihood (see Section~\ref{sec:synthetic_example}). The bias is plotted for varying sample sizes $N$ over the true EIG. The expectations of the EIG estimators are approximated using $10^4$ replications. 
    }
    \label{fig:bias-eig}
\end{figure}

The objective in this work is to select the optimal design $\xib$ from a finite, unordered set $\Xi$. We are not directly interested in approximating the EIG accurately for each $\xib \in \Xi$, but rather in identifying which design yields the highest EIG. Consequently, we focus on estimating the difference between two expected information gains:
\begin{eq}
\Delta (\xib_a, \xib_b) \defeq \EIG(\xib_a) - \EIG(\xib_b), 
\end{eq}
and confidently identifying the sign of the difference.
To estimate $\Delta(\xib_a, \xib_b)$, we employ a Monte Carlo estimator of the EIG. In particular, we use the ANMC estimator due to its favorable variance and bias properties, while noting that the methodology described in the following is agnostic to the specific estimator.\\
Further, we employ common random numbers (CRN) \cite{glasserman_guidelines_1992, rubinstein_simulation_2016} across all designs $\xib$ to reduce the variance of the estimator. To be precise, we use the same samples $\Theta$ from the prior for both designs $\xib_a$ and $\xib_b$. Using ANMC, this yields the following estimator for the EIG difference:
\begin{eq}
\widehat{\Delta}(\xib_a, \xib_b) &= \frac{1}{N} \sum_{i=1}^{N} \underbrace{\widehat{\mathrm{PMI}}^{\mathrm{RB}}_i(\Theta, \xib_a) -  \widehat{\mathrm{PMI}}^{\mathrm{RB}}_i(\Theta, \xib_b)}_{=:\delta_i(\xib_a, \xib_b)},
\end{eq}
where $\delta_i(\xib_a, \xib_b)$ denotes the difference between the i-th PMI estimates for designs $\xib_a$ and $\xib_b$, evaluated using the same prior samples $\Theta$.
To assess the confidence in the sign of the estimated EIG difference $\widehat{\Delta}(\xib_a, \xib_b)$, we employ bootstrap sampling \cite{efron_introduction_1994}. We generate $B$ bootstrap replicates by resampling the index set $\{1, \dots, N\}$ with replacement. For each bootstrap replicate $b = 1, \dots, N_B$, we draw indices $I_1^{*(b)}, \dots, I_N^{*(b)}$ independently and uniformly from $\{1, \dots, N\}$, and compute the bootstrap means:
\begin{eq}
\widehat{\Delta}^{*(b)}(\xib_a, \xib_b) = \frac{1}{N} \sum_{k=1}^{N} \delta_{I_k^{*(b)}}(\xib_a, \xib_b).
\label{eqn:bootstap_means}
\end{eq}
The bootstrap estimate of the probability that the EIG difference is positive is then given by:
\begin{eq}
\widehat{\mathbb{P}}\rbracket{\Delta(\xib_a, \xib_b) > 0} = \frac{1}{N_B} \sum_{b=1}^{N_B} \mathbf{1}\cbracket{ \widehat{\Delta}^{*(b)}(\xib_a, \xib_b) > 0 },
\label{eqn:diff_prob}
\end{eq}
where $\mathbf{1}\{\cdot\}$ denotes the indicator function, which equals 1 if the condition inside the braces is true and 0 otherwise.
Thus, $\widehat{\mathbb{P}}\rbracket{\Delta(\xib_a, \xib_b) > 0}$ is the proportion of bootstrap replicates for which the estimated EIG difference is positive and serves as a measure of confidence that design $\xib_a$ is more informative than design $\xib_b$. \\
We note that, due to the limited number of inner-loop samples, the estimator of the EIG difference $\widehat{\Delta}$ is generally biased.
Empirically, we observe that in most settings the bias of the NMC, LRNMC, and ANMC estimators of the EIG increases with the true EIG, thereby amplifying differences between competing designs. This behavior is consistent with the concavity of the logarithm and the observation that experiments with higher EIG induce greater variability in likelihood terms across prior samples, thereby increasing the Jensen gap.
In other words, the bias typically amplifies the EIG difference, making it easier to make a confident decision. In contrast, the RNMC estimator typically shows negative bias that decreases with increasing EIG. To put it differently, the RNMC estimator's bias typically reduces the EIG difference, which is problematic for optimal design decisions. \\
Figure~\ref{fig:bias-eig} shows the bias of the RNMC and the ANMC estimator of the EIG for a problem with a lognormal prior and likelihood (see Section~\ref{sec:synthetic_example}).
The expectations of the EIG estimators are approximated using $10^4$ independent replications. It can be seen that the magnitude of the bias reduces as expected with the number of samples $N$. The bias of the ANMC estimator is strictly positive and increases monotonically and superlinearly with increasing true EIG. On the other hand, the bias of the RNMC estimator is strictly negative and decreases monotonically with increasing EIG. This empirical behavior confirms that ANMC (whose bias coincides with that of NMC and LRNMC) tends to exaggerate differences between designs, facilitating decision-making, whereas RNMC compresses these differences, making it harder to reliably identify the optimal design. 

\subsection{Design Optimization}

A straightforward approach to BOED on a finite design set $\Xi$ is to draw a fixed number of parameter samples for each design independently, i.e., to generate $\Theta_k=\{\thetab_{k,i}\}_{i=1}^N$ with $\thetab_{k,i}\sim p(\thetab)$ for every $\xib_k\in\Xi$, evaluate the forward model $\model(\thetab_{k,i},\xib_k)$, and compute an EIG estimate $\widehat{\EIG}(\xib_k)$ for each design separately. The design with the largest estimated EIG is then selected. As an optional variance-reduction enhancement, this strategy can also be implemented using CRN, in which case a single shared sample set $\Theta$ is reused across all designs. 
However, regardless of whether independent samples or CRN are used, the method evaluates all designs with the same computational effort and provides no guidance on how many samples are needed to reliably distinguish between designs. In the numerical examples, we include this static design choice (SDC) approach as a baseline. \\
To overcome these limitations, we propose an adaptive design elimination (ADE) strategy that focuses computational effort on promising designs. 
Based on the aforementioned techniques, the core idea is to iteratively eliminate suboptimal designs from the candidate set $\Xi$ using a progressively larger sample set $\Theta$, and consequently, more model evaluations only for promising designs. 
At each iteration, we augment the parameter set $\Theta$ by sampling $\nnew$ new points from the prior distribution $p(\thetab)$ to compute an estimate of the EIG for all remaining designs. The design with the highest estimated EIG is selected as the current reference $\xib_{k^*}$. For each remaining design, we estimate the probability that its EIG is lower than that of $\xib_{k^*}$ using \eqref{eqn:diff_prob}. If this probability exceeds a threshold $q$, the design is marked for elimination. For increased robustness, a design is removed only if it consistently satisfies the elimination criterion for $N_C$ iterations in a row. The process continues with an increasing sample set size of $\Theta$ until only one design remains, which is then selected as the most informative. \\
As $\nnew$, $q$, and $N_C$ increase, the algorithm becomes more reliable in choosing the optimal design, but this also increases the number of model evaluations. Throughout all experiments, we use $\nnew=8$, $q = 0.99$, $N_C = 3$, and $N_B = 10^5$ as default values for comparisons against SDC, as these settings provide very high reliability and efficiency across applications. Appendix~\ref{appendix:ade_hyper} details the impact of these hyperparameters on performance for all numerical examples. \\
The procedure of the algorithm in combination with the ANMC estimator is summarized in Algorithm~\ref{alg:adaptive_elimination}. We recommend ANMC as the default choice due to its lower variance and its bias characteristics. However, the adaptive elimination procedure itself is independent of the underlying EIG estimator. RNMC or LRNMC can be used in exactly the same way by simply replacing the computation of $\widehat{\mathrm{PMI}}^{\mathrm{RB}}_i(\Theta,\xib_k)$ with the corresponding PMI scores of the chosen estimator. In the numerical examples, we also run the ADE algorithm with RNMC and LRNMC to benchmark their performance against ANMC. \\
The high efficiency of the algorithm in terms of model evaluations stems from multiple factors: The reuse of outer-loop samples for estimating the marginal likelihood, Rao--Blackwellization and common random numbers between candidate designs for variance reduction, the facilitating bias characteristics, and the early elimination of suboptimal designs. 
\begin{algorithm}[h!]
\setstretch{1.5}
\caption{Adaptive Design Elimination with ANMC}
\label{alg:adaptive_elimination}
\begin{algorithmic}
\REQUIRE $\Xi$, $\nnew=8$, $q=0.99$, $N_C=3$, $N_B=10^5$, $N_\epsilon=100$
\STATE Initialize candidate set: $\Xi_{\text{cand}} \gets \Xi$
\STATE Initialize parameter set: $\Theta \gets \emptyset$
\STATE Initialize counters: $C^{(k)} \gets 0$ for all $\xib_k \in \Xi$
\WHILE{$|\Xi_{\text{cand}}| > 1$}
    \STATE Sample new parameters: $\Theta \gets \Theta \cup \{\thetab_i\}_{i=1}^{\nnew}, \quad \thetab_i \sim p(\thetab)$
    \STATE Compute PMI scores: $P^{(k)} = \cbracket{\widehat{\mathrm{PMI}}^{\mathrm{RB}}_i(\Theta, \xib_k) }_{i=1}^{|\Theta|}$ for all $\xib_k \in \Xi_{\text{cand}}$ using $N_\epsilon$ noise samples (see \eqref{eqn:anmc})
    \STATE Identify reference design: $k^* = \argmax_k \; \overline{P^{(k)}}$
    \FOR{each $\xib_k \in \Xi_{\text{cand}} \setminus \cbracket{\xib_{k^*}}$}
        \STATE Compute differences: $D = \cbracket{P^{(k^*)}_i - P^{(k)}_i}_{i=1}^{|\Theta|}$
        \STATE Estimate probability: $\widehat{p} = \widehat{\mathbb{P}}\rbracket{\Delta(\xib_{k^*}, \xib_k) > 0}$ from $D$ using $N_B$ bootstrap samples (see \eqref{eqn:bootstap_means}, \eqref{eqn:diff_prob})
        \IF{$\widehat{p} > q$}
            \STATE $C^{(k)} \gets C^{(k)} + 1$
        \ELSE
            \STATE $C^{(k)} \gets 0$
        \ENDIF
    \ENDFOR
    \STATE Eliminate designs: $\Xi_{\text{cand}} \gets \Xi_{\text{cand}} \setminus \cbracket{\xib_k \mid C^{(k)} = N_C}$
\ENDWHILE
\RETURN $\Xi_{\text{cand}}$
\end{algorithmic}
\end{algorithm}

\section{Numerical Examples}
\label{chap:examples}

In the following, we evaluate the performance of the different estimators and the ADE algorithm using four numerical examples. The first two examples are adapted from the literature and, while not involving expensive computational models, provide valuable and well-established test cases. The first example considers a synthetic test case with an analytical solution for the EIG, while the second addresses the calibration of a pharmacokinetic model. Subsequently, we examine the calibration of a human lung model and, finally, the calibration of viscoplastic material parameters of a specimen.
In all examples, we use the default values $\nnew = 8$, $q = 0.99$, $N_C = 3$, and $N_B = 10^5$ for the ADE algorithm, and $N_\epsilon = 100$ for the ANMC estimator. For the NMC estimator, we set the number of inner samples such that $N = M^2$ to achieve optimal mean-squared error rates \cite{rainforth_nesting_2018}.

\subsection{Synthetic Test Case with Analytical Solution}
\label{sec:synthetic_example}

\begin{figure}[h!]
    \centering
    \includegraphics[width=\linewidth]{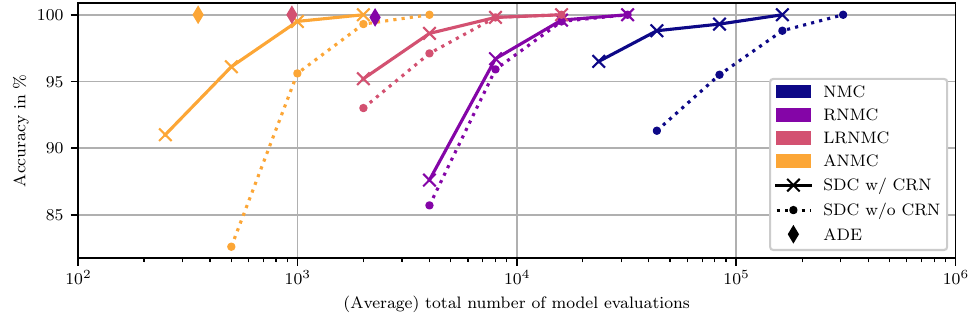}
    \caption{SDC (with and without CRN) and ADE performance over 1000 independent runs with the different estimators for the synthetic test case. The plot shows the fraction of runs in which the optimal design is correctly identified as a function 
    of the total number of model evaluations. In the case of ADE, the total number of model evaluations is averaged over runs.}
    \label{fig:sdc_ade_synthetic}
\end{figure}

To validate the proposed algorithm under controlled conditions, we consider a synthetic BOED problem for which the EIG admits a closed-form expression. This example is adapted from \cite{goda_unbiased_2022} and modified to feature a high-dimensional parameter space and a finite set of experimental designs. 
The prior of the unknown model parameters $\thetab \in \mathbb{R}^{100}_{+}$ is defined as a lognormal distribution:
\begin{eq}
    \log(\thetab) \sim \normal{\boldsymbol{0}, \sigma_p^2 I},
\end{eq}
where $\sigma_p^2=1$ controls the prior uncertainty. The likelihood model is Gaussian in log-space and depends on the design variable $\xi$:
\begin{eq}
    \log(\yb) \mid \thetab, \xi &\sim \normal{\model(\thetab, \xi),\; \sigma_n^2 I}, \qquad \xi \in \{1,2,\dots,10\}, \\
    \model(\thetab, \xi)&=A(\xi) \log \thetab.
\end{eq}
Here, $\yb \in \mathbb{R}^{50}_+$ denotes the observation vector and $\sigma_n^2=25$ is the observation noise variance. Each design $\xi$ corresponds to a dense linear transformation $A(\xi) \in \mathbb{R}^{50 \times 100}$, with entries generated once as $A_{ij}(\xi) = u_{ij,\xi} \cos(\xi)$, where $u_{ij,\xi} \sim \mathcal{U}(0,1)$. \\
The EIG for design $\xi$ admits a closed-form expression (see Appendix~\ref{appendix:synthetic}):
\begin{eq}
    \EIG(\xi) = \frac{1}{2} \log \det \rbracket{I + \frac{\sigma_p^2}{\sigma_n^2} A(\xi)^\top A(\xi)}.
\end{eq}
The EIG values for all designs are shown in Table~\ref{tab:synthetic-eig}, where the optimal design $\xi=3$ is highlighted. Note the small gap between the top two designs ($\xi=3$ and $\xi=6$), making this a challenging discrimination problem, especially in combination with the high-dimensional parameter space.
\begin{table}[h!]
\centering
\begin{tabular}{|c|c|c|c|c|c|c|c|c|c|c|}
\hline
$\xi$ & 1 & 2 & \cellcolor{cyellow}3 & 4 & 5 & 6 & 7 & 8 & 9 & 10 \\
\hline
$\mathrm{EIG}(\xi)$ & 3.610 & 2.504 & \cellcolor{cyellow}8.485 & 4.677 & 1.456 & 8.082 & 5.726 & 0.540 & 7.554 & 6.729 \\
\hline
\end{tabular}
\vspace{1mm}
\caption{Closed-form EIG values for the synthetic test case.  
The optimal design $\xi=3$ is highlighted.}
\label{tab:synthetic-eig}
\end{table} \\
Figure~\ref{fig:sdc_ade_synthetic} reports the accuracy of the SDC and the ADE approach in combination with the different estimators. The plot shows the fraction of 1000 independent runs that correctly identify the optimal design as a function of the total number of model evaluations. With the SDC algorithm, all designs are evaluated using the same number of model evaluations, whereas with ADE, the number of model evaluations varies across designs and runs. In this case, the average total number of model evaluations is reported. As expected, the accuracy increases with increasing number of model evaluations with the SDC algorithm for all estimators. Further, there is a significant performance improvement from NMC to RNMC, driven by sample reuse. Furthermore, moving from RNMC to LRNMC yields additional gains due to the improved bias characteristics. The transition from LRNMC to ANMC yields further improvement, attributable to the variance reduction achieved through Rao--Blackwellization. Overall, we can reduce the necessary number of model evaluations needed to reach 100\% accuracy by about two orders of magnitude from NMC to ANMC, and by about one order of magnitude from RNMC to ANMC. Moreover, the use of CRN significantly improves the performance of LRNMC, and especially of ANMC and NMC. Importantly, with the SDC approach, we would not know a priori how many model evaluations we would need to reliably obtain the most informative design. \\
This issue is addressed by the ADE algorithm, which automatically selects the required number of model evaluations. The algorithm selects the correct design in all runs for ANMC and LRNMC, and in 99.8\% of the runs for RNMC. The lower accuracy of RNMC can be attributed to the bias characteristics. Still, to achieve such high accuracy, the number of model evaluations is drastically reduced when comparing the ADE to the SDC algorithm with the same estimators, because ADE eliminates unpromising designs early. To put it in another perspective, we can reduce the necessary number of model evaluations needed to reach 100\% accuracy by about two orders of magnitude from SDC with RNMC to ADE with ANMC.
 \\

\subsection{Calibration of a Pharmacokinetic Model}
\label{sec:pk_example}

\begin{figure}[h!]
    \centering
    \includegraphics[width=\linewidth]{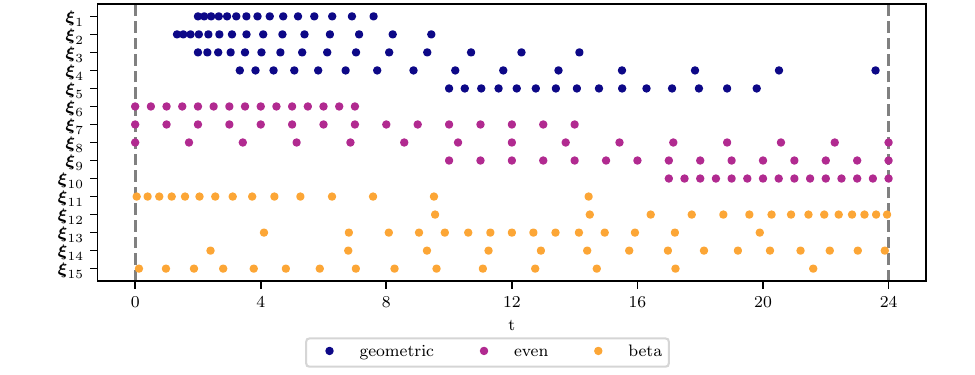}
    \caption{Sampling schedules for the pharmacokinetic model. For each scheme (geometric, uniform, beta), five designs with 15 sampling times in $[0,24]$ are generated. Rows correspond to the individual designs, and the points indicate the observation times.}
    \label{fig:pk_designs}
\end{figure}

We consider an example from \cite{ryan_towards_2014, zhang_scalable_2021} based on a one-compartment pharmacokinetic model, which describes the time evolution of drug concentration in the bloodstream following a single oral dose under first-order absorption and elimination dynamics. The drug concentration is observed at 15 blood sampling times $\xib \in [0,24]^{15}$ and is given by
\begin{eq}
\model(\thetab, \xib) 
&= \frac{FD}{V}\,\frac{k_a}{k_a - k_e}
\rbracket{ e^{-k_e \xib} - e^{-k_a \xib} },
\quad \thetab = [k_a, k_e, V], \\
\yb 
&= \model(\thetab, \xib)\,(1 + \epsb_m) + \epsb_a,
\end{eq}
where $k_a$ and $k_e$ denote the absorption and elimination rate constants, respectively, $V$ is the volume of distribution, $F$ represents the bioavailability, and $D$ the administered dose, assumed to be $400$.
The unknown parameters $\thetab$ follow a log-normal prior
\begin{eq}
\log k_a \sim \normal{\log(1), 0.05}, \quad \log k_e \sim \normal{\log(0.1), 0.05}, \quad \log V \sim \normal{\log(20), 0.05},
\end{eq}
with the constraint $k_a > k_e$. The measurement noise consists of independent multiplicative and additive components,
\begin{eq}
\epsb_m \sim \normal{\mathbf{0}, 0.01 I}, 
\qquad 
\epsb_a \sim \normal{\mathbf{0}, 0.1 I}.
\end{eq}
Following \cite{ryan_towards_2014}, we consider three parameterizations of the sampling schedule: a geometric, a uniform, and a beta scheme. For each scheme, we generate five  designs (see Figure~\ref{fig:pk_designs}). We approximate the true EIG values of all designs using the RNMC and the LRNMC estimator with $N=10^6$ samples, as shown in Table~\ref{tab:pk-eig}. Given the large number of samples, this provides a reliable estimate of the true EIG, allowing us to confidently identify the optimal design $\xib_{15}$.

\begin{figure}[h!]
    \centering
    \includegraphics[width=\linewidth]{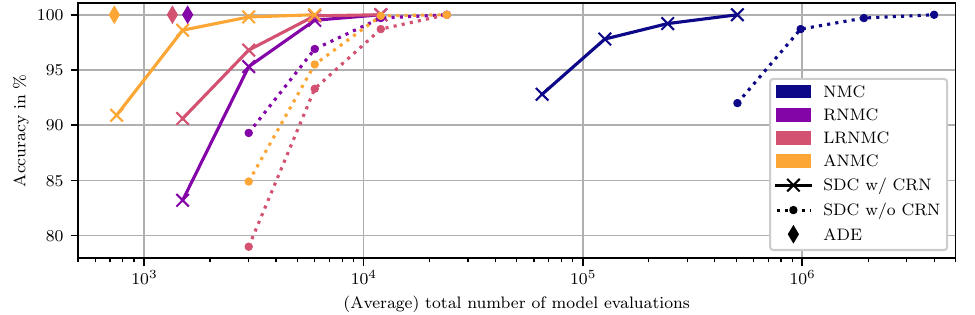}
    \caption{SDC (with and without CRN) and ADE performance over 1000 independent runs with the different estimators for the pharmacokinetic model example. The plot shows the fraction of runs in which the optimal design is correctly identified as a function 
    of the total number of model evaluations. In the case of ADE, the total number of model evaluations is averaged over runs.}
    \label{fig:sdc_ade_pk}
\end{figure}

\begin{table}[h!]
\centering
\begin{tabular}{|c|c|c|c|c|c|c|c|c|c|c|c|c|c|c|c|}
\hline
$\xib$ & $\xib_{1}$ & $\xib_{2}$ & $\xib_{3}$ & $\xib_{4}$ & $\xib_{5}$ & $\xib_{6}$ & $\xib_{7}$ & $\xib_{8}$ & $\xib_{9}$ & $\xib_{10}$ & $\xib_{11}$ & $\xib_{12}$ & $\xib_{13}$ & $\xib_{14}$ & \cellcolor{cyellow}$\xib_{15}$ \\
\hline
$\widehat{\mathrm{EIG}}_{\mathrm{RNMC}}(\xib)$ & 2.909 & 3.308 & 3.457 & 3.519 & 2.945 & 3.269 & 3.650 & 3.698 & 3.040 & 2.603 & 3.856 & 2.914 & 3.226 & 3.498 & \cellcolor{cyellow}4.083 \\
\hline
$\widehat{\mathrm{EIG}}_{\mathrm{LRNMC}}(\xib)$ & 2.909 & 3.308 & 3.457 & 3.519 & 2.945 & 3.270 & 3.650 & 3.698 & 3.040 & 2.603 & 3.856 & 2.914 & 3.227 & 3.499 & \cellcolor{cyellow}4.084 \\
\hline
\end{tabular}
\vspace{1mm}
\caption{Approximated EIG values for the pharmacokinetic model calibration example using $N=10^6$ samples.  The optimal design $\xib_{15}$ is highlighted.}
\label{tab:pk-eig}
\end{table} 
Figure~\ref{fig:sdc_ade_pk} reports the accuracy of the SDC and ADE approaches in combination with the different estimators for this example. The plot shows the fraction of 1000 independent runs that correctly identify the optimal design as a function of the (average) total number of model evaluations. As in the previous example, the SDC algorithm allocates the same number of model evaluations to all designs, whereas for ADE the number of model evaluations varies across designs and runs, and the average total number is reported.
Consistent with the previous example, the use of CRN for the SDC algorithm leads to performance improvements across all estimators, with particularly noticeable gains for NMC and ANMC. The largest improvements, however, stem from sample reuse. Without CRN, RNMC slightly outperforms both ANMC and LRNMC, presumably due to its lower estimator variance. When CRN is enabled, the variance is further reduced and the impact of the bias characteristics becomes dominant, resulting in the expected ordering of estimator performance. In particular, we observe improvements from NMC to RNMC due to sample reuse, followed by further gains from RNMC to LRNMC driven by the improved bias characteristics, and from LRNMC to ANMC as a result of additional variance reduction through Rao--Blackwellization.
As before, a key limitation of the SDC approach is that the number of model evaluations required to reliably identify the optimal design is not known a priori. This issue is addressed by the ADE algorithm, which automatically determines an appropriate computational budget while eliminating unpromising designs early. As shown in Figure~\ref{fig:sdc_ade_pk}, ADE achieves 100\% accuracy across all estimators while requiring significantly fewer model evaluations than SDC to achieve this accuracy. With the ADE approach, the expected performance ordering from RNMC to LRNMC to ANMC is visible. The computational effort to achieve such high accuracy can be reduced by more than one order of magnitude when comparing SDC with the RNMC estimator to ADE with the ANMC estimator.

\subsection{Calibration of a Human Lung Model}

\begin{figure}[b!]
    \centering
    \includegraphics[width=0.29\linewidth]{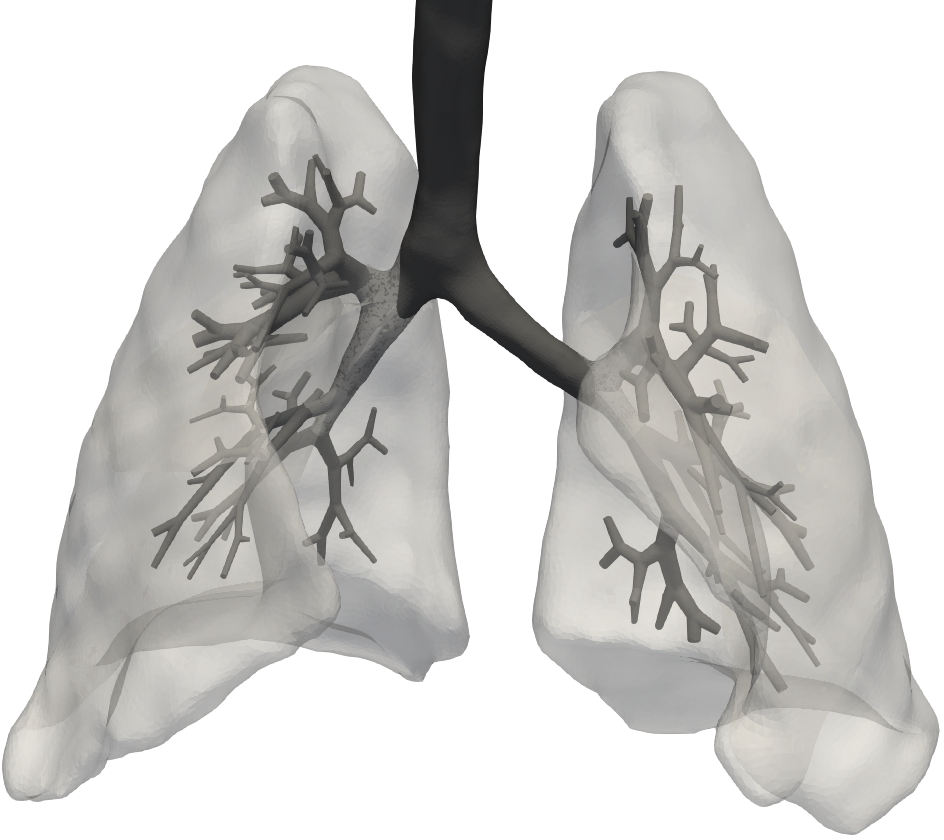} \hspace{2cm}
    \includegraphics[width=0.3\linewidth]{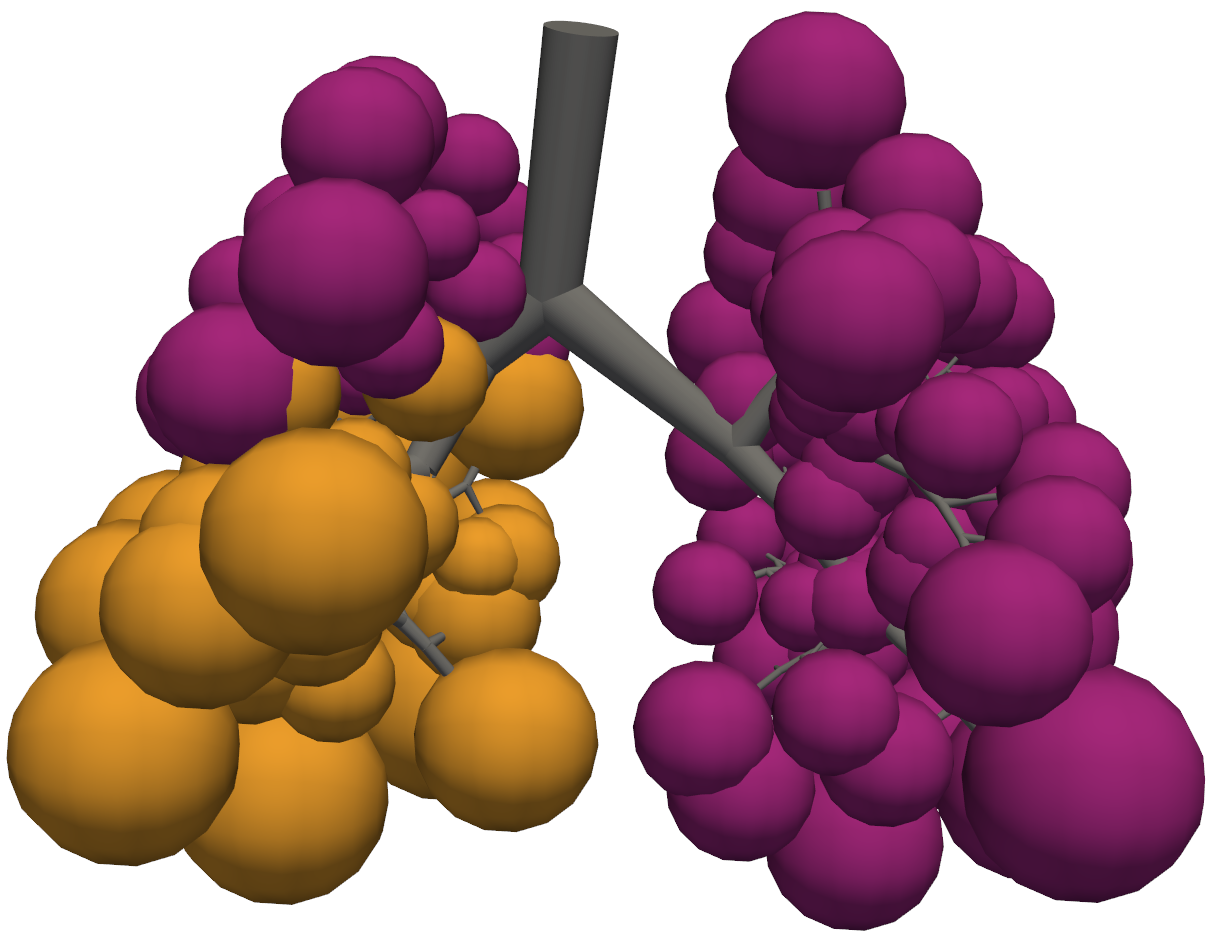}
    \caption{Segmented human lung (left) that is used for the generation of the computational model (right). The airways of the computational model are visualized in grey, and the terminal units as spheres. The two colors indicate the two distinct regions used to parameterize the terminal units.}
    \label{fig:lung_geometry}
\end{figure}

\begin{figure}[t!]
    \centering
    \includegraphics[width=\linewidth]{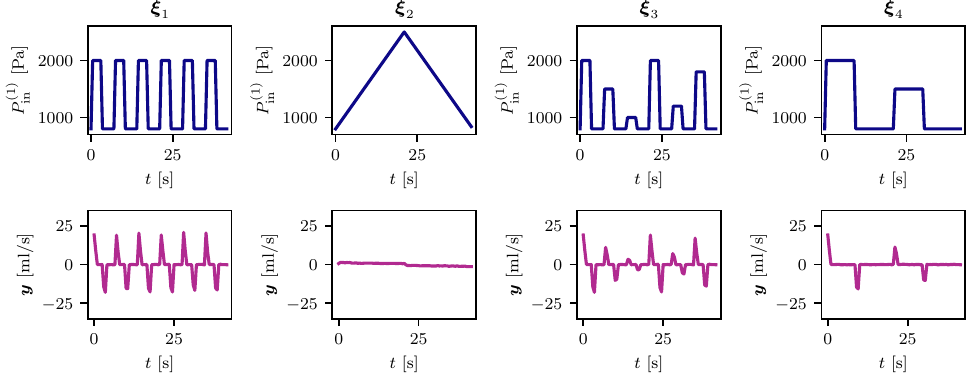}
    \caption{Ventilation pressure protocols of the four experimental designs are shown in the top row, and the corresponding measurements $\yb$ for a random realization of the parameters $\thetab$ and the noise $\epsb_a$ and $\epsb_m$ are shown in the bottom row.}
    \label{fig:lung_maneuvers}
\end{figure}

\begin{figure}[b!]
    \centering
    \includegraphics[width=\linewidth]{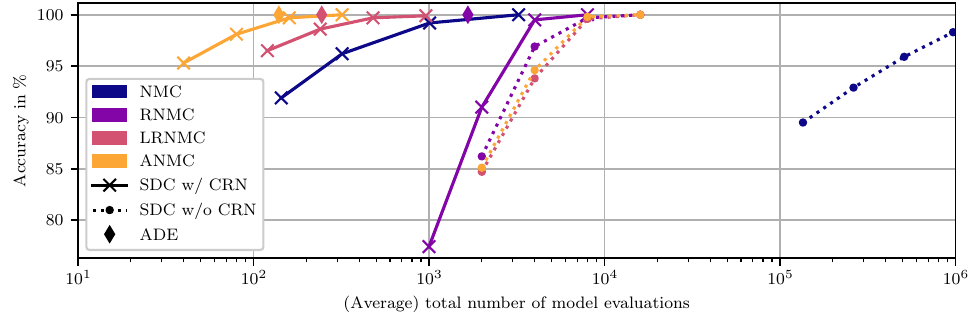}
    \caption{SDC (with and without CRN) and ADE performance over 1000 runs with the different estimators for the lung model example. The plot shows the fraction of runs in which the optimal design is correctly identified as a function 
    of the total number of model evaluations. In the case of ADE, the total number of model evaluations is averaged over runs.}
    \label{fig:sdc_ade_lung}
\end{figure}

In this example, we consider the Bayesian calibration of a reduced-dimensional human lung model that captures the respiratory mechanics of a mechanically ventilated patient. For some background on such lung models, the reader is referred to \cite{ismail_coupled_2013, roth_comprehensive_2017, geitner_pressure-_2024, rixner_patient-specific_2025, sablewski_predicting_2026}. \\
The central airways of the lungs are described by an idealized pipe-flow model, which can be simplified under suitable assumptions to the following equations:
\begin{eq}
	C \dfrac{\mathrm{d}}{\mathrm{d}t}\rbracket{ \dfrac{1}{2} \rbracket{ P_{\mathrm{in}} + P_{\mathrm{out}} } - \widetilde{P}_{\mathrm{ext}}}  + Q_{\mathrm{out}} - Q_{\mathrm{in}} + C\cdot R_{\mathrm{visc}} \dfrac{\mathrm{d}}{\mathrm{d}t} \rbracket{ Q_{\mathrm{out}} - Q_{\mathrm{in}}} = 0,   \\ 
    \dfrac{I}{2} \dfrac{\mathrm{d}}{\mathrm{d}t} \rbracket{Q_{\mathrm{in}} + Q_{\mathrm{out}}} + \dfrac{1}{2} \rbracket{ R_{\mathrm{\mu}} + R_{\mathrm{conv}}} \cdot \textcolor{black}{\rbracket{ Q_{\mathrm{in}} + Q_{\mathrm{out}} }}  + P_{\mathrm{out}} - P_{\mathrm{in}} = 0, \label{eq:0D_pipe2}
 \end{eq}
where adjacent airways are coupled through the interface flow rates $Q_{\mathrm{in}}$, $Q_{\mathrm{out}}$ and pressures $P_{\mathrm{in}}$, $P_{\mathrm{out}}$. The parameters $C$, $I$, and $R$ determine capacitive, inductive, and resistive effects, respectively. Here, we resolve the first eight generations of the airway tree with these equations. The remaining local lung compartments, which comprise the peripheral airways and parenchyma, are modeled as so-called terminal units, which are described by a Kelvin-Voigt model with nonlinear, Ogden-type elasticity:
 \begin{eq}
    P_{\mathrm{TU}} - P_{\mathrm{pl}} = \frac{\kappa}{\beta} \rbracket{ \frac{V_{\mathrm{TU}}}{V_{\mathrm{TU,0}}} }^{-1} \rbracket{ 1 - \rbracket{ \frac{V_{\mathrm{TU}}}{V_{\mathrm{TU,0}}}}^{-\beta}} + \eta  \dfrac{\mathrm{d}}{\mathrm{d}t} \rbracket{\frac{V_{\mathrm{TU}}}{V_{\mathrm{TU,0}}} },
\end{eq}
where $P_{\mathrm{TU}}$ is the pressure within the terminal unit and $P_{\mathrm{pl}}$ is the surrounding pleural pressure. 
The ratio $\frac{V_{\mathrm{TU}}}{V_{\mathrm{TU,0}}}$ denotes the terminal unit’s instantaneous volume relative to its initial, stress‑free reference volume $V_{\mathrm{TU,0}}$. The parameters $\kappa$ and $\beta$ determine the stiffness and curvature of the pressure--volume relation, respectively, while $\eta$ scales viscous effects. \\
We assume the lung exhibits heterogeneous material parameters due to pathological changes and aim to calibrate the parameters $\kappa$, $\beta$, and $\eta$. We divide the lung into two regions with distinct parameters based on the lung lobes (see Figure~\ref{fig:lung_geometry}). In total, this gives us six calibration parameters $\thetab = [\kappa_1, \beta_1, \eta_1, \kappa_2, \beta_2, \eta_2]$  with independent uniform priors: 
\begin{eq}
    \kappa_1, \kappa_2 \sim \mathcal{U}(1000, 5000) \, [\text{Pa}], \quad \eta_1, \eta_2 \sim \mathcal{U}(100, 500) \, [\text{Pa} \cdot \text{s}], \quad \beta_1, \beta_2 \sim \mathcal{U}(-8, -4).
\end{eq}
As experimental designs, we consider four different pressure-controlled ventilation protocols similar to those used in clinical practice. In the simulation, this represents the pressure that is applied at the inlet of the first airway. We monitor inlet flow at the first airway $Q_{\mathrm{in}}^{(1)}$ over time as a surrogate for the clinically observed flow at the ventilator. The measurements at the times $\boldsymbol{t}$ are subjected to additive Gaussian noise $\epsb_a$ with variance $\sigma_a^2=0.1^2 [\text{ml}/\text{s}]^2$ and multiplicative Gaussian noise $\epsb_m$ with variance $\sigma_m^2=0.05^2$:
\begin{eq}
    \yb &=  Q_{\mathrm{in}}^{(1)}(\thetab, \xib, \boldsymbol{t} )(1+\epsb_{m}) + \epsb_{a}, \quad \boldsymbol{t} = [t_1,\ldots,t_T] \\
    \epsb_{m} &\sim \normal{\boldsymbol{0}, \sigma_m^2 I}, \quad \epsb_{a} \sim \normal{\boldsymbol{0}, \sigma_a^2 I}.
\end{eq}
The relatively large multiplicative noise reflects the difficulty of accurately measuring flow rates in practice. \\
Figure~\ref{fig:lung_maneuvers} displays the applied ventilation pressure protocols in the top row and the corresponding measurements $\yb$ for a random realization of the parameters $\thetab$ and the noise $\epsb_a$ and $\epsb_m$ in the bottom row. \\
As a reference, we approximate the EIG of all four designs using RNMC and LRNMC with $N=10^6$ (see Table~\ref{tab:lung-eig}). Due to the high number of samples, we can confidently identify design $\xib_1$ as the optimal one, closely followed by design $\xib_3$. 
\label{sec:lung_example}
\begin{table}[h!]
\centering
\begin{tabular}{|c|c|c|c|c|}
\hline
$\xib$ & \cellcolor{cyellow}$\xib_{1}$ & $\xib_{2}$ & $\xib_{3}$ & $\xib_{4}$ \\
\hline
$\widehat{\mathrm{EIG}}_{\mathrm{RNMC}}(\xib)$ & \cellcolor{cyellow}7.339 & 4.469 & 7.125 & 5.434 \\
\hline
$\widehat{\mathrm{EIG}}_{\mathrm{LRNMC}}(\xib)$ & \cellcolor{cyellow}7.356 & 4.470 & 7.137 & 5.436 \\
\hline
\end{tabular}
\vspace{1mm}
\caption{Approximated EIG values for the human lung calibration example using $N=10^6$ samples.  The optimal design $\xib_1$ is highlighted.}
\label{tab:lung-eig}
\end{table} \\
Figure~\ref{fig:sdc_ade_lung} reports the accuracy of the SDC and ADE approaches in combination with the different estimators for finding the optimal design. As in the previous examples, the plot shows the fraction of 1000 runs that correctly identify the optimal design as a function of the (average) total number of model evaluations. Due to the increased computational cost of the model, we precompute a set of $10^6$ input samples and the corresponding model outputs for each design. In each run, we randomly draw samples from this precomputed set, without duplicates within a run. Since some samples might be used in more than one run, the runs are not fully independent. \\
Regarding the SDC results, the observed behavior is comparable to that in the previous examples. The use of CRN substantially improves performance for LRNMC, ANMC, and NMC, and also yields a modest improvement for RNMC. In the absence of CRN, the performance of RNMC, LRNMC, and ANMC is largely comparable, with similar computational effort required to achieve a given level of accuracy. Without CRN, NMC performs particularly poorly, requiring substantially more model evaluations than the other estimators to achieve high accuracy.
Interestingly, NMC with CRN outperforms RNMC, which can plausibly be attributed to the unfavorable bias characteristics of RNMC. Nevertheless, the expected ordering among the remaining estimators is recovered. LRNMC outperforms NMC due to effective sample reuse, while ANMC yields further improvements over LRNMC through variance reduction via Rao--Blackwellization.\\
Again, the SDC approach does not provide guidance on the number of model evaluations required to reliably identify the optimal design. This limitation is addressed by the ADE algorithm, which adaptively allocates computational effort and eliminates unpromising designs early. Using ADE, the correct design is identified in all runs for ANMC, LRNMC, and RNMC.
For all estimators, ADE requires fewer model evaluations to achieve this level of accuracy than SDC with the respective estimators. The relative reduction is less pronounced than in the previous cases, since only four designs are considered here and, consequently, the potential for early elimination is more limited.
Still, to put the overall gains into perspective, the number of model evaluations needed to attain 100\% accuracy can be reduced by approximately two orders of magnitude when moving from SDC with RNMC without CRN to ADE with ANMC.

\subsection{Calibration of a Viscoplastic Material Model}
\label{sec:visco_example}
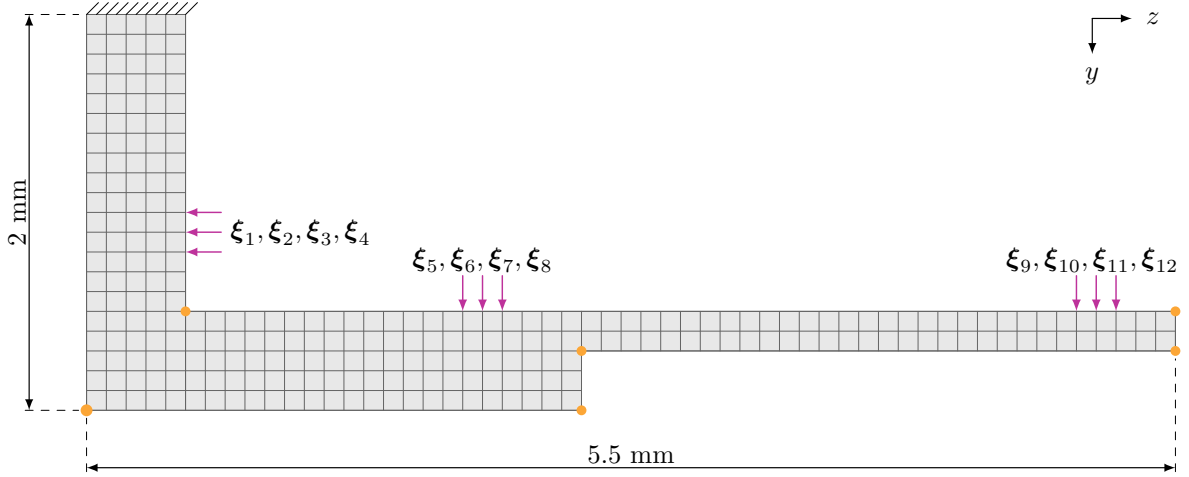
\begin{figure}[t!]
    \centering
\pgfmathsetlengthmacro{\figunit}{\textwidth/63}

\begin{tikzpicture}[
    x=\figunit,
    y=\figunit,
    line cap=round,
    line join=round,
    >={Stealth[length=1.6ex]}
]

\path[use as bounding box] (-6.5,-3.5) rectangle (56.5,21.5);

\def\H{20}      
\def\L{55}      
\def\Wv{5}      
\def\Hb{5}      
\def\Hr{2}      
\def\Xs{25}     

\definecolor{meshgray}{RGB}{100,100,100}
\definecolor{fillgray}{RGB}{232,232,232}
\definecolor{loadmagenta}{RGB}{190,50,155}
\definecolor{ptorange}{RGB}{252,166,54}


\fill[fillgray] (0,0) rectangle (\Xs,\Hb);

\fill[fillgray] (0,0) rectangle (\Wv,\Hb);

\fill[fillgray] (0,\Hb) rectangle (\Wv,\H);

\fill[fillgray] (\Xs,\Hb-\Hr) rectangle (\L,\Hb);


\foreach \x in {\Wv,\numexpr\Wv+1\relax,...,\Xs} {
    \draw[meshgray, line width=0.22pt] (\x,0) -- (\x,\Hb);
}
\foreach \y in {0,1,...,\Hb} {
    \draw[meshgray, line width=0.22pt] (\Wv,\y) -- (\Xs,\y);
}

\foreach \x in {0,1,...,\Wv} {
    \draw[meshgray, line width=0.22pt] (\x,0) -- (\x,\H);
}
\foreach \y in {0,1,...,\H} {
    \draw[meshgray, line width=0.22pt] (0,\y) -- (\Wv,\y);
}

\foreach \x in {\Xs,\numexpr\Xs+1\relax,...,\L} {
    \draw[meshgray, line width=0.22pt] (\x,\Hb-\Hr) -- (\x,\Hb);
}
\foreach \y in {\numexpr\Hb-\Hr\relax,\numexpr\Hb-\Hr+1\relax,\Hb} {
    \draw[meshgray, line width=0.22pt] (\Xs,\y) -- (\L,\y);
}

\draw[meshgray, line width=0.35pt]
    (0,0) --
    (\Xs,0) --
    (\Xs,\Hb-\Hr) --
    (\L,\Hb-\Hr) --
    (\L,\Hb) --
    (\Wv,\Hb) --
    (\Wv,\H) --
    (0,\H) -- cycle;

\draw[meshgray, line width=0.35pt] (0,\H) -- (\Wv,\H);
\foreach \x in {0,0.5,...,5} {
    \draw[black, line width=0.35pt] (\x,\H) -- ++(0.6,0.6);
}

\fill[ptorange] (0,0) circle (0.3);
\fill[ptorange] (\Wv,\Hb) circle (0.25);
\fill[ptorange] (\Xs,0) circle (0.25);
\fill[ptorange] (\Xs,\Hb-\Hr) circle (0.25);
\fill[ptorange] (\L,\Hb-\Hr) circle (0.25);
\fill[ptorange] (\L,\Hb) circle (0.25);


\draw[dashed, line width=0.3pt] (-3.1,0) -- (-0.4,0);
\draw[dashed, line width=0.3pt] (-3.1,\H) -- (-0.4,\H);
\draw[latex-latex, line width=0.45pt] (-2.9,0) -- (-2.9,\H);
\node[rotate=90] at (-3.5,10) {$2\ \mathrm{mm}$};

\draw[dashed, line width=0.3pt] (0,-3.1) -- (0,-0.4);
\draw[dashed, line width=0.3pt] (\L,-3.1) -- (\L,\Hb-\Hr-0.4);
\draw[latex-latex, line width=0.45pt] (0,-2.9) -- (\L,-2.9);
\node at (27.5,-2.25) {$5.5\ \mathrm{mm}$};


\foreach \yy in {8,9,10} {
    \draw[loadmagenta, line width=0.55pt, -latex] (\Wv+1.8,\yy) -- (\Wv,\yy);
}
\node[anchor=west] at (\Wv+1.8,9.0) {$\xib_1,\xib_2,\xib_3,\xib_4$};

\foreach \xx in {19, 20, 21} {
    \draw[loadmagenta, line width=0.55pt, -latex] (\xx,\Hb+1.8) -- (\xx,\Hb);
}
\node at (20.0,7.6) {$\xib_5,\xib_6,\xib_7,\xib_8$};

\foreach \xx in {50,51,52} {
    \draw[loadmagenta, line width=0.55pt, -latex] (\xx,\Hb+1.8) -- (\xx,\Hb);
}
\node at (50.8,7.6) {$\xib_9,\xib_{10},\xib_{11},\xib_{12}$};

\begin{scope}[shift={(50.8,19.8)}]
    \draw[-latex, line width=0.45pt] (0,0) -- (2.0,0);
    \draw[-latex, line width=0.45pt] (0,0) -- (0,-1.8);
    \node[anchor=west] at (2.25,0.02) {$z$};
    \node[anchor=north] at (0,-2.0) {$y$};
\end{scope}

\end{tikzpicture}
\caption{Geometry and finite element mesh of the specimen. The purple arrows indicate the different locations of the applied displacements for the respective designs, while the yellow dots denote the displacement observation points.}
\label{fig:visco_mesh}
\end{figure}

\begin{figure}[h!]
    \centering
    \includegraphics[width=\linewidth]{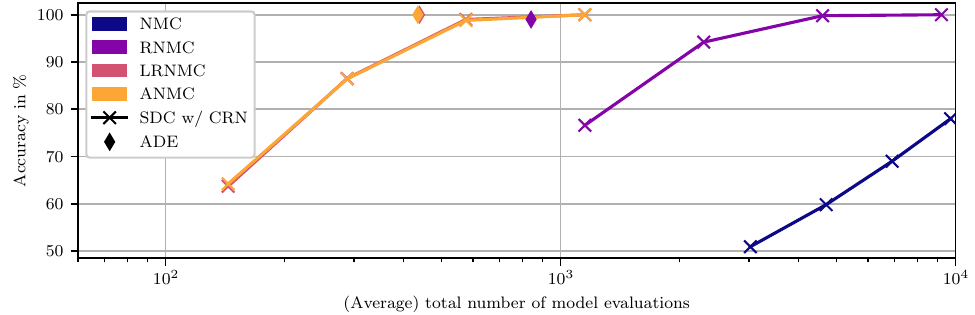}
    \caption{SDC (with CRN) and ADE performance over 1000 runs with the different estimators for the viscoplastic material model example. The plot shows the fraction of runs in which the optimal design is correctly identified as a function 
    of the total number of model evaluations. In the case of ADE, the total number of model evaluations is averaged over runs.}
    \label{fig:sdc_ade_visco}
\end{figure}

This study investigates the calibration of a material model for the viscoplastic behavior exhibited by a solid specimen under mechanical loading. 
To this end, we employ the finite strain viscoplasticity model used in \cite{ana_adaptive_2026}, which is based on a thermodynamically consistent reformulation of the Johnson--Cook model presented in \cite{mareau_thermodynamically_2020}.
Furthermore, the adaptive estimate interpolation algorithm developed in \cite{ana_adaptive_2026} is used for the constitutive update, as it has demonstrated significant efficiency and robustness improvements when applied to this material model. \\
The model is formulated based on the multiplicative decomposition
\begin{eq}
    \boldsymbol{F}  = \boldsymbol{F}_{\text{e}} \boldsymbol{F}_{\text{p}},
\end{eq}
where $\boldsymbol{F}_{\text{e}}$ and $\boldsymbol{F}_{\text{p}}$ denote the elastic and plastic deformation gradients, respectively.
This decomposition defines the right elastic Cauchy--Green deformation tensor $ \boldsymbol{C}_{\text{e}} = \boldsymbol{F}_{\text{e}}^T \boldsymbol{F}_{\text{e}}$ and the elastic Green--Lagrange strain tensor $\boldsymbol{E}_{\text{e}} = \frac{1}{2} \left( \boldsymbol{C}_{\text{e}} - \boldsymbol{I} \right)$. 
Assuming Saint Venant--Kirchhoff hyperelasticity, the elastic response is described using the Mandel stress
\begin{eq}
    \boldsymbol{M}_{\text{e}}  = \lambda \mathrm{tr}\left( \boldsymbol{E}_{\text{e}} \right) \boldsymbol{C}_{\text{e}} + 2 \mu \boldsymbol{C}_{\text{e}} \boldsymbol{E}_{\text{e}}.
\end{eq}
The Lamé constants are related to the Young's modulus $E$ and the Poisson's ratio $\nu$ through
\begin{eq}
	\lambda  = \frac{\nu  E}{\left( 1 + \nu \right)  \left( 1 - 2  \nu \right)}, \quad
	\mu      = \frac{E}{2  \left( 1 + \nu \right)}.
\end{eq}
Furthermore, the plastic strain rate is specified as
\begin{eq}	
\dot{\varepsilon}_{\text{p}} = \begin{cases}
		\dot{\varepsilon}_{\text{p}, 0}   \exp \left[ \frac{1}{C}  \left( \frac{\overline{\sigma}}{\sigma_{\text{yield}}(\varepsilon_{\text{p}})} - 1 \right) \right]  - \dot{\varepsilon}_{\text{p}, 0}, & \text{for } \overline{\sigma} \ge \sigma_{\text{yield}}(\varepsilon_{\text{p}}), \\
		0, & \text{otherwise}.
	\end{cases}
\end{eq}
Here, the equivalent stress is defined as the von Mises stress with $ \overline{\sigma} = \sqrt{\frac{3}{2} \boldsymbol{M}_{\text{e}}': \boldsymbol{M}_{\text{e}}'}$, where $\boldsymbol{M}_{\text{e}}'$ specifies the deviatoric part of the elastic Mandel stress. \\
In the plastic flow rule, $\dot{\varepsilon}_{\mathrm{p},0}$ is a reference plastic strain rate, while $C$ governs the strain rate sensitivity. 
Moreover, the yield stress exhibits isotropic hardening according to
\begin{eq}
    \sigma_{\text{yield}} = \sigma_{\text{yield}, 0} + H \varepsilon_{\text{p}}^M,
\end{eq}
where $\sigma_{\mathrm{yield},0}$ denotes the initial yield stress, while $H$ and $M$ represent the hardening modulus and exponent, respectively. \\
We solve the solid mechanics problem on the geometry depicted in Figure~\ref{fig:visco_mesh} using the finite element method. 
The forward model is implemented in the open-source code \texttt{4C}~\cite{4c_4c_nodate}, while the scheduling and orchestration of the simulations are carried out using \texttt{QUEENS}~\cite{biehler_queens_2025}. 
The mesh consists of $260$ quadratic hexahedral elements with $3621$ nodes.
Inertial effects are incorporated, and the mass density is fixed to $\rho = 7850\,\mathrm{kg/m^3}$, representative of structural steel. 
Accordingly, the vector of unknown material parameters is restricted to the elasticity and viscoplasticity parameters with
\begin{eq}
\thetab = [E, \nu, \dot{\varepsilon}_{\mathrm{p},0}, C, \sigma_{\mathrm{yield},0}, H, M],
\end{eq}
and we assume independent uniform priors:
\begin{eq}
&E \sim \mathcal{U}(50, 300)\,[\text{GPa}], \quad
\nu \sim \mathcal{U}(0.1, 0.5), \quad
\dot{\varepsilon}_{\mathrm{p},0} \sim \mathcal{U}(0.001, 1)\,[1/\text{s}], \quad
C \sim \mathcal{U}(0.1, 1.0), \\
&\sigma_{\mathrm{yield},0} \sim \mathcal{U}(500, 2000)\,[\text{MPa}], \quad
H \sim \mathcal{U}(100, 2000)\,[\text{MPa}], \quad
M \sim \mathcal{U}(0.1, 0.5).
\end{eq}
The specimen is loaded through prescribed time-dependent displacements, and the experimental designs correspond to different loading configurations, defined by the displacement location and displacement rate. 
We consider four constant displacement rates
\begin{eq}
0.1~\mathrm{[mm/s]} \; (\xib_1, \xib_5, \xib_9), \quad
0.2~\mathrm{[mm/s]} \; (\xib_2, \xib_6, \xib_{10}), \quad
0.4~\mathrm{[mm/s]} \; (\xib_3, \xib_7, \xib_{11}), \quad
1.0~\mathrm{[mm/s]} \; (\xib_4, \xib_8, \xib_{12}).
\end{eq}
and three loading positions (see Figure~\ref{fig:visco_mesh}), resulting in 12 experimental designs.
The model output consists of the displacement components in $y$- and $z$-direction measured at six spatial locations $\xb_j$, $j=1,\dots,6$, and discrete times $t_i$. The maximum prescribed displacement of $0.2$ [mm] is identical across designs, and we assume a fixed temporal observation resolution of 50 [1/s] across all experiments. Consequently, designs with higher displacement rates are observed over a shorter time horizon and therefore yield fewer measurements (e.g., a doubling of the displacement rate results in half as many observations). The observations are given by
\begin{eq}
y_{k,i,j}
=
u_k(\thetab, \xib, t_i, \xb_j) + \epsilon_{k,i,j},
\qquad k \in \{y,z\},
\end{eq}
where $u_k(\thetab, \xib, t_i, \xb_j)$ denotes the model-predicted displacement component in direction $k$. The measurement noise follows a Student-$t$ distribution
\begin{eq}
\epsilon_{k,i,j}
\sim t_{\nu}(0,s).
\end{eq}
This choice demonstrates the applicability of the proposed methodology to non-Gaussian observation models while providing a representation of occasional outlier measurements. We choose $\nu=10$ degrees of freedom and a scale parameter $s=0.002$ [mm]. \\
As a reference, we approximate the EIG of all designs using RNMC and ANMC with $N=10^3$ samples. The resulting values identify $\xib_9$ as the most informative loading configuration for parameter calibration. Interestingly, although higher displacement rates are expected to enhance the influence of viscoplastic effects and therefore increase the information content of individual measurements, the results indicate that the larger number of observations obtained at lower displacement rates has a stronger overall impact. Consequently, a design with the lowest displacement rate and the corresponding longest observation horizon achieves the highest EIG.
\begin{table}[h!]
\centering
\begin{tabular}{|c|c|c|c|c|c|c|c|c|c|c|c|c|}
\hline
$\xib$ & $\xib_{1}$ & $\xib_{2}$ & $\xib_{3}$ & $\xib_{4}$ & $\xib_{5}$ & $\xib_{6}$ & $\xib_{7}$ & $\xib_{8}$ & \cellcolor{cyellow}$\xib_{9}$ & $\xib_{10}$ & $\xib_{11}$ & $\xib_{12}$ \\
\hline
$\widehat{\mathrm{EIG}}_{\mathrm{RNMC}}(\xib)$ & 1.099 & 0.889 & 0.774 & 0.549 & 6.421 & 6.058 & 5.654 & 4.914 & \cellcolor{cyellow}6.662 & 6.342 & 5.836 & 5.353 \\
\hline
$\widehat{\mathrm{EIG}}_{\mathrm{ANMC}}(\xib)$ & 1.171 & 0.952 & 0.775 & 0.588 & 11.678 & 8.947 & 7.158 & 5.697 & \cellcolor{cyellow}26.185 & 17.681 & 12.598 & 8.700 \\
\hline
\end{tabular}
\vspace{1mm}
\caption{Approximated EIG values for the viscoplastic material model calibration example using $N=10^3$ samples.  The optimal design $\xib_9$ is highlighted.}
\label{tab:visco-eig}
\end{table} \\
Figure~\ref{fig:sdc_ade_visco} compares the performance of SDC and ADE across the different estimators in identifying the optimal design. As in the previous examples, the results are reported as the fraction of 1000 runs that correctly recover the optimal design versus the (average) number of model evaluations.
Due to the high computational cost of the forward model, we precompute a set of $10^3$ input samples and the corresponding model outputs for each design. In each run, samples are drawn from this set without replacement, introducing slight dependencies between runs when samples are reused across different runs.
Results for SDC without CRN are omitted, as achieving acceptable accuracy would require an infeasible number of model evaluations. \\
For SDC with CRN, we observe the expected performance improvements when moving from NMC to RNMC. Moreover, transitioning from RNMC to LRNMC reduces the computational effort by nearly an order of magnitude for comparable accuracy, due to its favorable bias properties. ANMC and LRNMC exhibit nearly identical performance, which can be explained by the large number of observations, since the variance contribution from the observation noise becomes negligible compared to the variance induced by the parameter samples. \\
Using ADE, the optimal design is identified in all runs for ANMC and LRNMC, and in 99\% of the runs for RNMC. Across all estimators, ADE requires fewer model evaluations than SDC with the corresponding estimator to achieve the same level of accuracy, owing to the early elimination of unpromising designs. To achieve the high accuracy attained by ADE, SDC with CRN requires approximately 2.5 times more model evaluations when using ANMC or LRNMC, and about four times more when using RNMC. \\
Overall, this example demonstrates that the proposed methodology can efficiently and reliably identify optimal experimental designs even for computationally expensive finite element models and in the presence of non-Gaussian, design-dependent observation models.

\section{Conclusion}
\label{chap:conclusion}

We present an efficient and reliable approach for selecting optimal experimental designs from finite candidate sets, tailored to applications involving expensive computational models. The method combines several complementary ideas to drastically reduce the number of required forward model evaluations while maintaining high reliability. \\
At its core, the approach relies on nested Monte Carlo estimation of the expected information gain, enhanced through systematic reuse of outer-loop samples \cite{huan_accelerated_2010, huan_simulation-based_2013}. Importantly, we exclude the current outer-loop sample from the inner loop, which improves the bias behavior when comparing competing designs. Additional variance reduction is achieved through common random numbers \cite{glasserman_guidelines_1992, rubinstein_simulation_2016} across designs and Rao--Blackwellization \cite{robert_monte_2004}, enabling more reliable comparisons at lower sample sizes. \\
To address the challenge of selecting an appropriate sample size, we introduce an adaptive design elimination algorithm. The method progressively increases the sample size while eliminating inferior designs based on bootstrap estimates of the probability that one design outperforms another. Requiring consistent elimination decisions over multiple iterations ensures robustness. Moreover, by focusing computational effort on promising candidates, the algorithm significantly reduces the overall cost. \\
Across four challenging numerical examples, we demonstrate that the proposed framework reduces the number of model evaluations by up to several orders of magnitude compared to standard approaches, while consistently identifying the optimal design. The method performs robustly across diverse settings, including high-dimensional parameter spaces, complex noise models, and computationally intensive and complex forward models. Although the impact of the individual methodological components varies across problems, their complementary strengths result in consistently reliable and efficient design selection. We also provide practical default hyperparameter choices that perform reliably across applications and analyze their impact in Appendix~\ref{appendix:ade_hyper}. \\
Future work could extend the proposed framework to goal-oriented Bayesian experimental design \cite{zhong_goal-oriented_2026, bernardo_expected_1979} or settings with nuisance parameters \cite{bartuska_small-noise_2022, feng_layered_2019}.

\section*{Acknowledgments}
MD and WAW gratefully acknowledge financial support by BREATHE, an ERC-2020-ADG Project, Grant Agreement ID 101021526. DCA and WAW gratefully acknowledge support by the Bavarian Ministry of Economic Affairs, Regional Development and Energy [project ”Industrialisierbarkeit Festkörperelektrolyte”].

\section*{CRediT authorship contribution statement}
\textbf{Maximilian Dinkel}: Conceptualization, Methodology, Software, Validation, Formal analysis, Investigation, Data Curation, Writing -- Original Draft, Writing -- Review \& Editing, Visualization. 
\textbf{Dragos C. Ana}: Software, Validation, Investigation, Writing -- Review \& Editing.
\textbf{Benedikt Goderbauer}: Software, Visualization, Writing -- Review \& Editing.
\textbf{Wolfgang A. Wall}: Conceptualization, Resources, Methodology, Writing -- review \& editing, Supervision, Project administration, Funding acquisition.

\section*{Declaration of competing interests}
The authors declare that they have no known competing financial interests or personal relationships that could have appeared to
influence the work reported in this paper.

\section*{Data availability}
The code and data used to generate the results in this work are publicly available at: \url{https://github.com/maxdinkel/boed}.

\section*{Declaration of generative AI and AI-assisted technologies in the manuscript preparation process.}
During the preparation of this work, the authors used Microsoft 365 Copilot (AI models based on OpenAI's GPT-4 and GPT-5-class systems) to assist with drafting, language, and clarity of the manuscript, as well as simple coding tasks. The authors reviewed and edited the output as needed and take full responsibility for the content of the published article.

\printbibliography

\appendix

\section{Derivation of Expected Information Gain}

\label{appendix:eig}
As shown in \cite{rainforth_modern_2024}, the expected information gain can be derived as:
\begin{eq}
    \EIG(\xib) 
    &= \E_{p(\yb \mid \xib)} \sbracket{ \IG(\xib, \yb) } \\
    &= \E_{p(\yb \mid \xib)} \sbracket{ \mathrm{H}[p(\thetab)] - \mathrm{H}[p(\thetab \mid \yb, \xib)] } \\
    &=  \int p(\yb \mid \xib) \sbracket{p(\thetab \mid \yb, \xib) \log p(\thetab \mid \yb, \xib) - p(\thetab) \log p(\thetab) } d\thetab d\yb \\
    &= \int p(\thetab) p(\yb \mid \thetab, \xib)  \sbracket{ \log p(\thetab \mid \yb, \xib) - \log p(\thetab) }  d\yb d\thetab \\
    &=  \int p(\thetab) p(\yb \mid \thetab, \xib)  \sbracket{ \log p(\yb \mid \thetab, \xib) - \log p(\yb \mid \xib)}  d\yb d\thetab.
\end{eq}

\section{Derivation of Closed-Form EIG of Synthetic Test Case}
\label{appendix:synthetic}

Since mutual information is invariant under homeomorphisms \cite{kraskov_estimating_2004} and the EIG equals the mutual information \cite{rainforth_modern_2024}, it follows that the EIG is invariant under the logarithmic reparameterization, which defines a homeomorphism on $\mathbb{R}_+$. 
Under this transformation, the model in \ref{sec:synthetic_example} becomes a Gaussian linear model:
\begin{eq}
    \thetab' &\sim \normal{\boldsymbol{0}, \sigma_p^2 I}, \quad &\thetab' &= \log(\thetab),   \\
    \yb' \mid \thetab', \xi &\sim 
    \normal{A(\xi)\thetab', \sigma_n^2 I}, \quad &\yb' &= \log(\yb).
\end{eq}
By conjugacy of the Gaussian prior and likelihood, the posterior distribution is also Gaussian \cite{bishop_pattern_2006, murphy_machine_2012}:
\begin{eq}
    \thetab' \mid \yb', \xi &\sim 
    \normal{\mub_{\mathrm{post}}, \Sigma_{\mathrm{post}}}, \\
    \mub_{\mathrm{post}} &= \Sigma_{\mathrm{post}} \frac{1}{\sigma_n^2} A(\xi)^\top \yb', \\
    \Sigma_{\mathrm{post}}^{-1} &= \frac{1}{\sigma_p^2} I + \frac{1}{\sigma_n^2} A(\xi)^\top A(\xi) .
\end{eq}
For Gaussians, we can compute the entropy (see e.g. \cite{murphy_machine_2012}) and consequently the information gain in closed form:
\begin{eq}
    \IG(\xib, \yb') &= \frac{1}{2} \log \det(\sigma_p^2 I) - \frac{1}{2} \log \det(\Sigma_{\mathrm{post}}) \\
    &=\frac{1}{2} \log \det(\sigma_p^2 I) + \frac{1}{2} \log \det \rbracket{\frac{1}{\sigma_p^2} I + \frac{1}{\sigma_n^2} A(\xi)^\top A(\xi)} \\
    &= \frac{1}{2} \log \det\!\left(I + \frac{\sigma_p^2}{\sigma_n^2} A(\xi)^\top A(\xi)\right).
\end{eq}
Since the information gain in the linear Gaussian model is independent of the realization of $\mathbf{y}'$, it follows that
\begin{eq}
    \EIG(\xib) = \IG(\xib, \mathbf{y}').
\end{eq}

\section{Hyperparameter Study of ADE algorithm}
\label{appendix:ade_hyper}
\begin{figure}[h!]
    \centering
    \includegraphics[width=\linewidth]{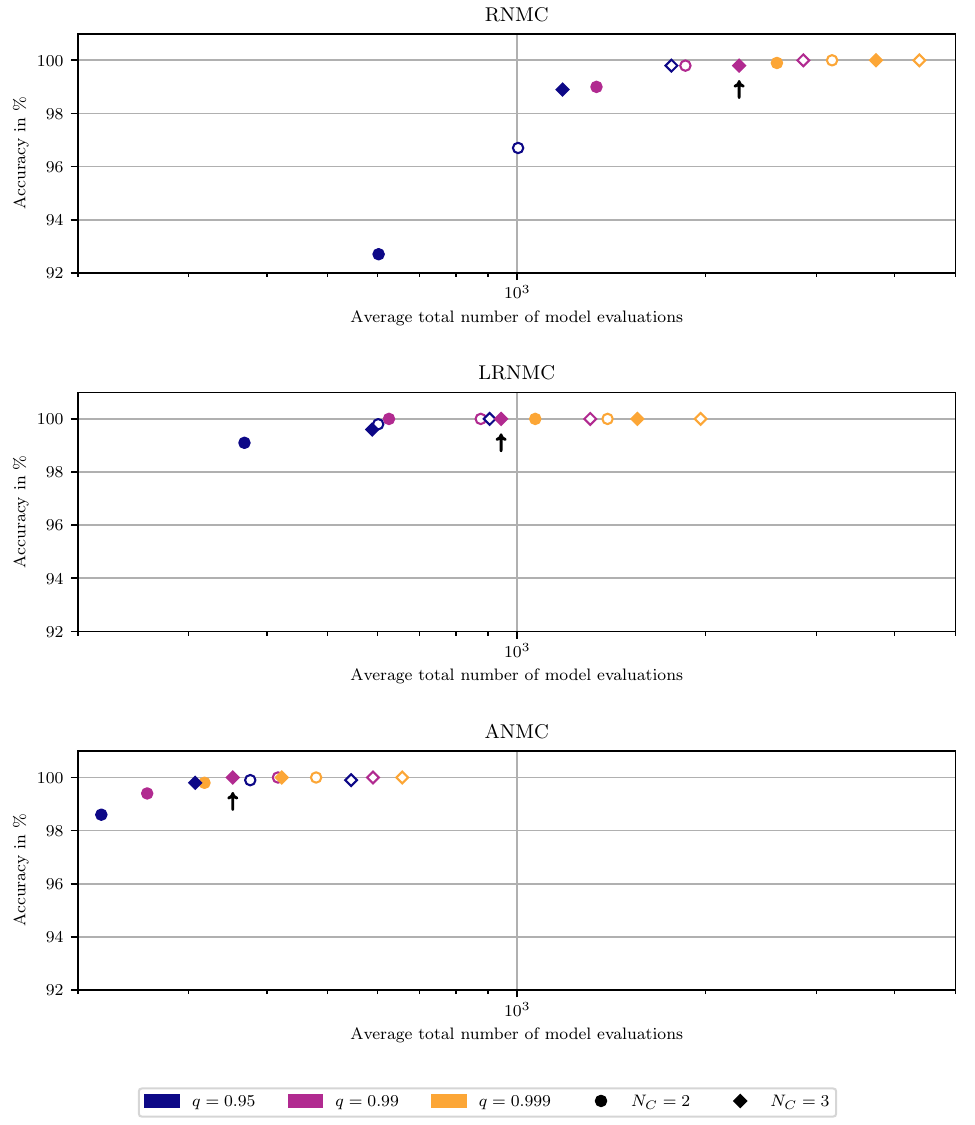}
    \caption{ADE performance over 1000 independent runs with the different estimators and with varying hyperparameter settings for the synthetic test case example (see Section~\ref{sec:synthetic_example}). The plot shows the fraction of runs in which the optimal design is correctly identified as a function 
    of the average total number of model evaluations. Filled markers correspond to $\nnew = 8$, while open markers indicate $\nnew = 16$. The default hyperparameter configuration ($q = 0.99$, $N_C = 3$, $\nnew = 8$) is highlighted by a black arrow.}
    \label{fig:ade_synthetic}
\end{figure} 
\begin{figure}[h!]
    \centering
    \includegraphics[width=\linewidth]{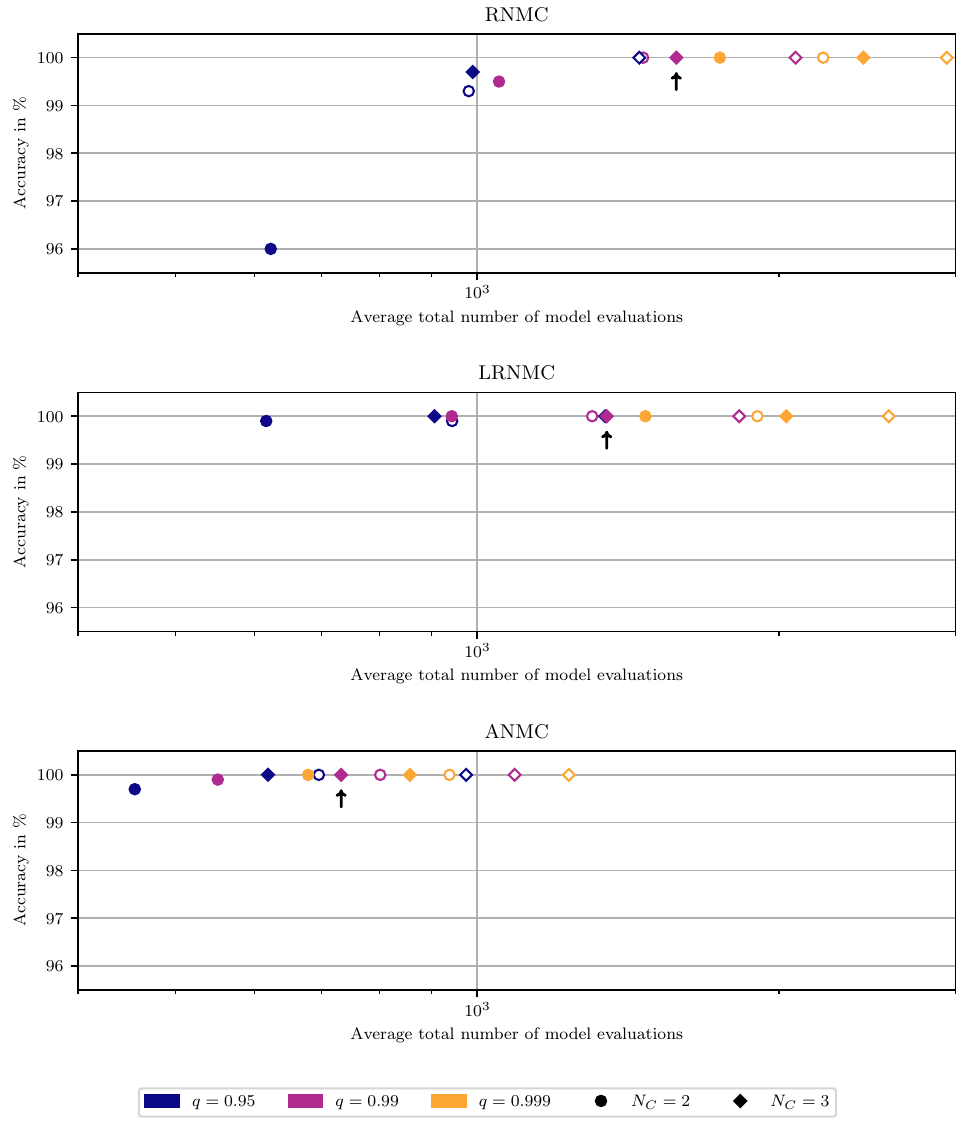}
    \caption{ADE performance over 1000 independent runs with the different estimators and with varying hyperparameter settings for the pharmacokinetic model example (see Section~\ref{sec:pk_example}). The plot shows the fraction of runs in which the optimal design is correctly identified as a function 
    of the average total number of model evaluations. Filled markers correspond to $\nnew = 8$, while open markers indicate $\nnew = 16$. The default hyperparameter configuration ($q = 0.99$, $N_C = 3$, $\nnew = 8$) is highlighted by a black arrow.}
    \label{fig:ade_pk}
\end{figure} 
\begin{figure}[h!]
    \centering
    \includegraphics[width=\linewidth]{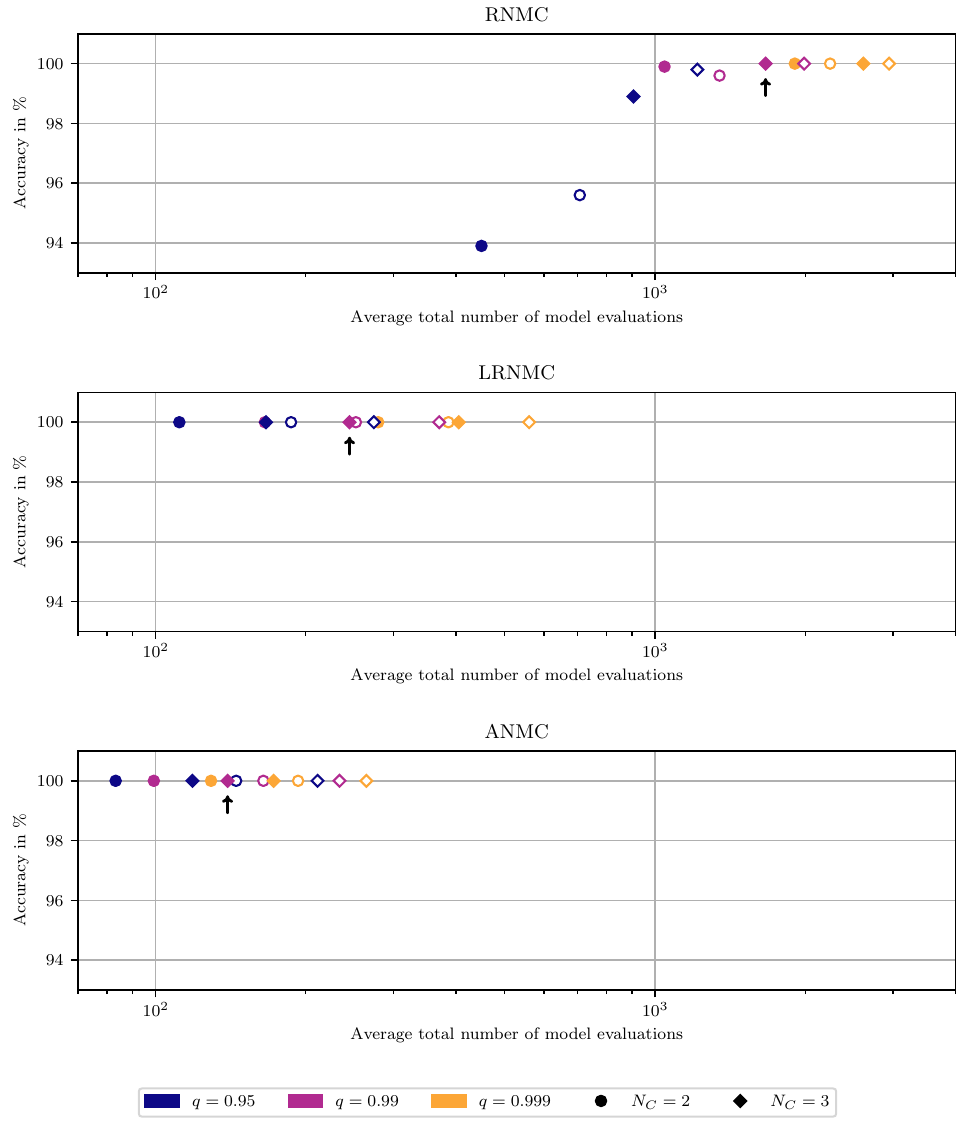}
    \caption{ADE performance over 1000 runs with the different estimators and with varying hyperparameter settings for the lung model example (see Section~\ref{sec:lung_example}). The plot shows the fraction of runs in which the optimal design is correctly identified as a function 
    of the average total number of model evaluations. Filled markers correspond to $\nnew = 8$, while open markers indicate $\nnew = 16$. The default hyperparameter configuration ($q = 0.99$, $N_C = 3$, $\nnew = 8$) is highlighted by a black arrow.}
    \label{fig:ade_lung}
\end{figure} 
\begin{figure}[h!]
    \centering
    \includegraphics[width=\linewidth]{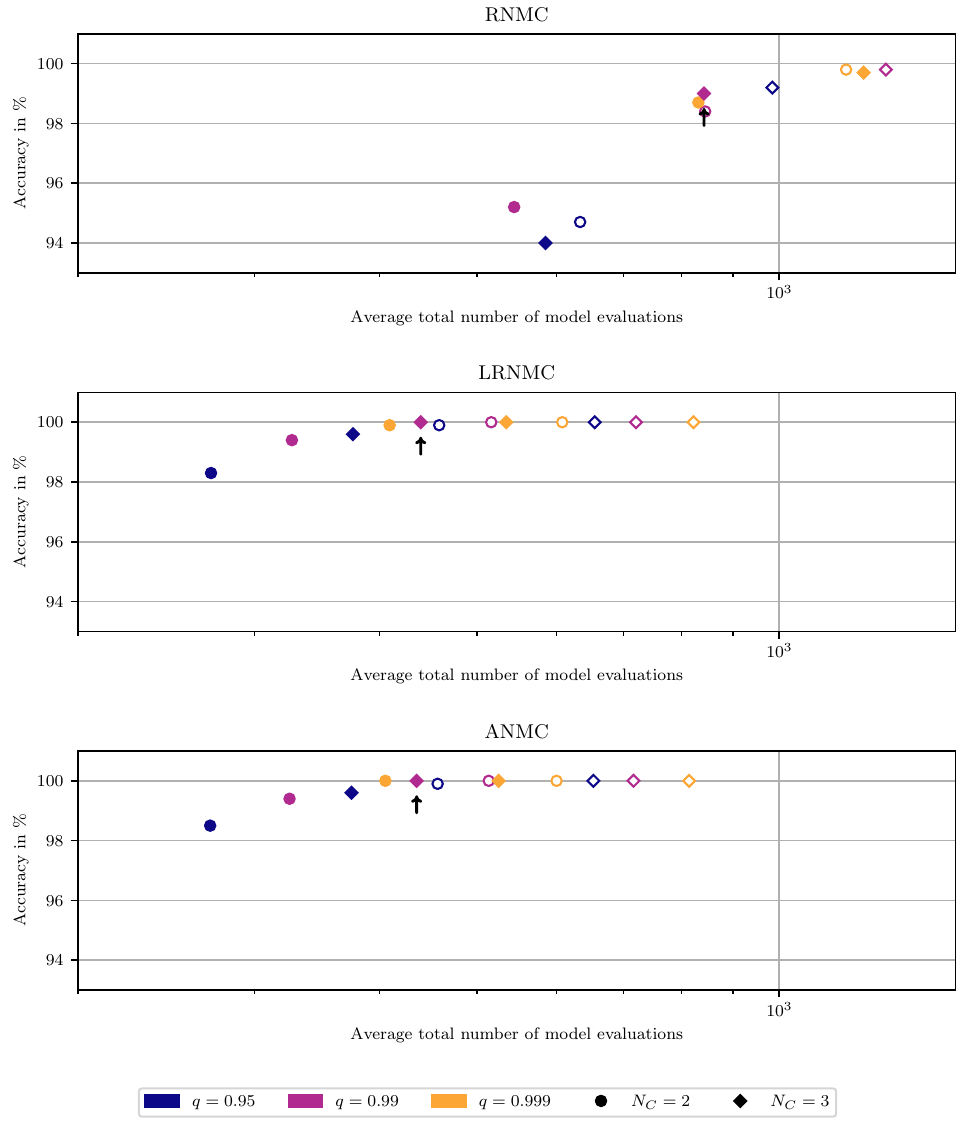}
    \caption{ADE performance over 1000 runs with the different estimators and with varying hyperparameter settings for the viscoplastic material model example (see Section~\ref{sec:visco_example}). The plot shows the fraction of runs in which the optimal design is correctly identified as a function 
    of the average total number of model evaluations. Filled markers correspond to $\nnew = 8$, while open markers indicate $\nnew = 16$. The default hyperparameter configuration ($q = 0.99$, $N_C = 3$, $\nnew = 8$) is highlighted by a black arrow.}
    \label{fig:ade_visco}
\end{figure} 
We examine how the three key ADE hyperparameters affect performance when combined with the RNMC, ANMC, and LRNMC estimators. The plots show the fraction of 1000 runs that correctly identify the optimal design versus the average number of model evaluations. We vary the probability threshold $q \in \{0.95, 0.99, 0.999\}$ (indicated by the color), the elimination threshold $N_C \in \{2, 3\}$ (indicated by the marker), and the number of new samples per iteration $\nnew \in \{8, 16\}$ (indicated by the marker filling). The default setting is highlighted by a black arrow. \\
Across all examples from Chapter~\ref{chap:examples}, increasing $q$, $N_C$, and $\nnew$ raises computational cost and typically improves accuracy. The default choice $q=0.99$, $N_C=3$, $\nnew=8$ provides a good trade-off, achieving 100\% accuracy with LRNMC and ANMC in all tested examples.

\end{document}